**Atomic-scale manipulation and *in situ* characterization with scanning tunneling microscopy**

*Wonhee Ko, Chuanxu Ma, Giang D. Nguyen, Marek Kolmer, and An-Ping Li\**

Dr. W. Ko, Dr. C. Ma, Dr. M. Kolmer, Dr. A.-P. Li
Center for Nanophase Materials Sciences, Oak Ridge National Laboratory, Oak Ridge,
Tennessee 37831, USA

Dr. G. D. Nguyen
Center for Nanophase Materials Sciences, Oak Ridge National Laboratory, Oak Ridge,
Tennessee 37831, USA
Stewart Blusson Quantum Matter Institute, University of British Columbia, Vancouver,
British Columbia V6T 1Z4, Canada

E-mail: apli@ornl.gov

Keywords: scanning tunneling microscopy, atomic-scale manipulation, *in situ*

Scanning tunneling microscope (STM) has presented a revolutionary methodology to the nanoscience and nanotechnology. It enables imaging the topography of surfaces, mapping the distribution of electronic density of states, and manipulating individual atoms and molecules,

all at the atomic resolution. In particular, the atom manipulation capability has evolved from

fabricating individual nanostructures towards the scalable production of the atomic-sized

devices bottom-up. The combination of precision synthesis and *in situ* characterization of the

atomically precise structures has enabled direct visualization of many quantum phenomena

and fast proof-of-principle testing of quantum device functions with real-time feedback to

guide the improved synthesis. In this article, several representative examples are reviewed to

demonstrate the recent development of atomic scale manipulation. Especially, the review

focuses on the progress that address the quantum properties by design through the precise

control of the atomic structures in several technologically relevant materials systems. Besides

conventional STM manipulations and electronic structure characterization with single-probe

STM, integration of multiple atomically precisely controlled probes in a multiprobe STM

system vastly extends the capability of *in situ* characterization to a new dimension where the charge and spin transport behaviors can be examined from mesoscopic to atomic length scale. The automation of the atomic scale manipulation and the integration with the well-established lithographic processes would further push this bottom-up approach to a new level that combines reproducible fabrication, extraordinary programmability, and the ability to produce large-scale arrays of quantum structures.

## 1. Introduction

A few years after the invention of scanning tunneling microscope (STM),[1] it was discovered that atomic resolution and precise tip positioning of STM enables manipulation of individual atoms.[2] In 1990, Eigler et al. demonstrated the smallest letters written by Xe atoms on a metal surface with the atom manipulation (**Figure 1**). The work gained immediate attention and inspired many imaginations into the nano-world. As Toumey argued, rather than the Feynman's paper, the three most important scientific events leading to the National Nanotechnology Initiative were the invention of the STM, the invention of the atomic force microscope (AFM), and the first manipulation of atoms.[3] Especially, atomic scale manipulation of the materials allowed researchers to perform condensed matter experiments in an idealized situation where all atom positions are defined precisely. Controlled with ultra-high-vacuum (UHV) and cryogenic temperature, STM can characterize the quantum states of the atomic-scale structures in precisely defined environment.[4]

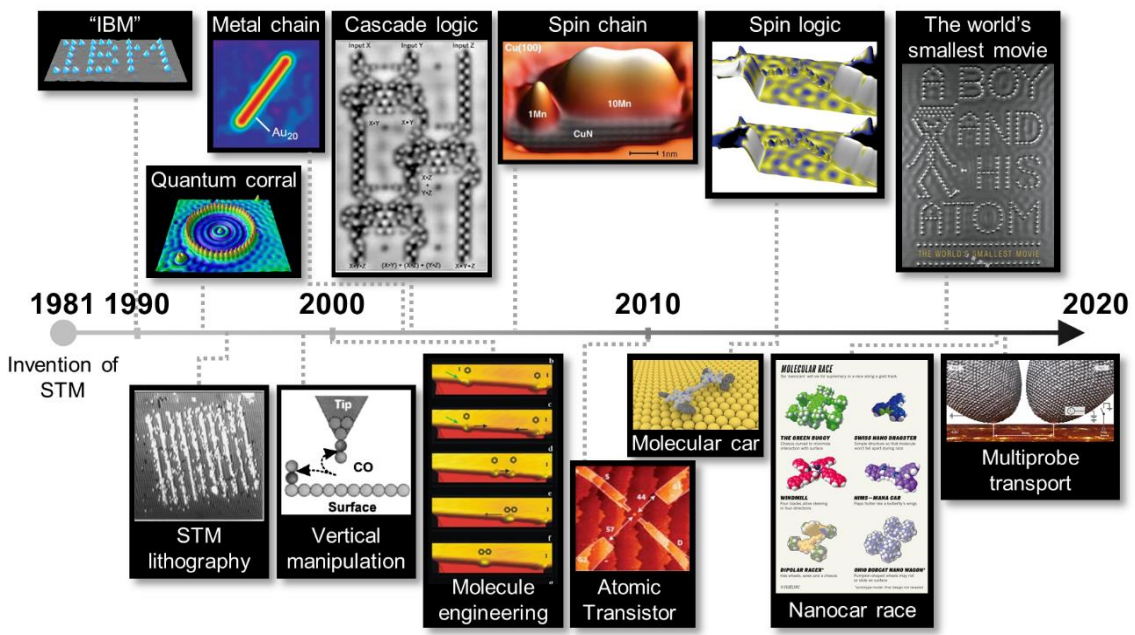

**Figure 1.** A historical view of the atomic scale manipulation with STM. Images from left to right: the word "IBM" written by Xe atoms (adapted with permission.[2] Copyright 1990,

Springer Nature. Image originally created by IBM Corporation); a circular quantum corral built from Fe atoms (adapted with permission.[4] Copyright 1993, AAAS. Image originally created by IBM Corporation); STM lithography demonstrated on hydrogenated Si(100) surface by tip-induced dehydrogenation (reprinted with permission.[5] Copyright 1994, AIP Publishing); vertical manipulation of single CO molecules (reprinted with permission.[6] Copyright 1998, American Physical Society); intra- and inter-molecular reactions induced by STM tip (adapted with permission.[7] Copyright 2000, American Physical Society); the gold atomic chain that exhibits 1D band structure (reprinted with permission.[8] Copyright 2002, AAAS); a logic gate composed of cascading CO molecules (reprinted with permission.[9] Copyright 2002, AAAS); a chain of Mn atoms that have net spin moment (reprinted with permission.[10] Copyright 2006, AAAS); transistor composed of a few phosphorus atoms on Si(100) (reprinted with permission.[11] Copyright 2010, Springer Nature); spin logic gate built from the chain of Fe atoms (adapted with permission.[12] Copyright 2011, AAAS); molecular car propelled by the tunneling current (reprinted with permission.[13] Copyright 2015, Springer Nature); the world's smallest movie built from atoms (adapted with permission.[14] Image courtesy of IBM Corporation, Copyright by IBM Corporation); Nanocar Race of molecular cars (reprinted with permission.[15] Copyright 2017, Springer Nature); multiprobe transport measurement on the artificial atomic wire (adapted with permission.[16] Copyright 2017, IOP Publishing. Image courtesy of Scienta Omicron GmbH, Germany).

Since the first demonstration by Eigler et al., significant progress was achieved in atomic scale manipulation with STM for both technical development and scientific understanding as illustrated in the historical snapshots in Figure 1. Understanding the physical mechanisms of interactions between the STM tip and the sample resulted in various manipulation methods. Traditionally, "atom manipulation" meant the STM techniques moving individual atoms and molecules physisorbed on the substrate surface with the STM tip. In the simplest way, it was achieved by bringing the STM tip close to the adsorbates to pull or push them to the new positions, named "lateral manipulation".[2] Then, Bartels et al. discovered that atoms or molecules could be transferred from the surface to the tip or vice versa with a proper bias voltage pulse. They applied the method to manipulate atoms by picking them up, moving to a new position, and then dropping them down, which was dubbed "vertical manipulation".[6, 17] Atomic scale manipulation with STM, however, is not limited to moving physisorbed atoms or molecules. Even in the 1990s, it was already shown that the electric field and tunneling electrons from the STM tip could induce a change in chemical bonds of materials in the atomic scale. Lyding et al. desorbed hydrogen atoms on the silicon surface in atomic precision,[5] and Hla et al. tailored molecules so that they underwent specific chemical reactions.[7] After these early discoveries, manipulation techniques were continuously developed to enhance their precision, reliability, and scalability. The first structure built by atom manipulation consists of 35 atoms,[2] but now atomically precise structures composed of thousands of atoms can be built with STM.[11, 18] Even a stop-motion video film was made from the images of manipulated atoms, named "A Boy and His Atom: The World's Smallest Movie".[14] Also, the size of the adsorbates available for manipulation increased from single atoms to macro-molecules that are composed of hundreds of atoms. Not only was the simple linear motion of macro-molecules demonstrated,[19] but the sophisticated motion of molecular machines was also accomplished by STM manipulation, exemplified by molecular cars moving along the predefined tracks with energies fueled by the tunneling electrons.[20] Development of STM hardware with advanced functions greatly expanded the applicability of atom manipulation. One prominent example is a multiprobe STM, which usually has four probes that can be operated independently and scanned with atomic resolution. The state-of-the-art multiprobe STM can manipulate atoms and molecules with all four probes simultaneously. The system was used in a celebrated event of Nanocar Race, where several molecular cars with different shapes were deposited on a gold surface and propelled simultaneously with four independent probes.[15] Another advantage of multiprobe STM is its ability to measure transport properties down to the atomic scale. Both atomic scale manipulation and *in situ* transport characterization can be performed successively without exposure to the ambient environment.[16]

The application of the atomic scale manipulation with STM relies on its ability to tailor the structure of the materials in atomic precision and tune their physical properties as designed. This was first demonstrated by Crommie et al in a seminal work on quantum corrals.[4] A circular corral was built by assembling individual atoms on a metal surface, which confined the surface electrons to form a textbook type of two-dimensional (2D) particle-in-a-box-like waveform. Since then, tunability developed vastly as demonstrated in the manipulation for various kinds of physical properties, including electronic, magnetic, chemical, and mechanical properties. Arranging atoms or molecules in specific geometries created electronic bands different from that of the substrate. Nilius et al. built a chain of metal atoms to display 1D band structure,[8] and Gomes et al. assembled a triangular lattice of carbon monoxide molecules to transform metal surface states to a 2D massless Dirac band.[21] Using spin-polarized STM for atom manipulations made it possible to manipulate and characterize magnetic nanostructures.[22] Heinrich group and Wiesendanger group showed various kinds of spin chains made of transition metal atoms, which displayed potentials of storing digital information and performing logic gate function.[10, 12, 23] Mechanical interactions between the molecules were exploited by Heinrich et al. to build molecular cascade, where carbon monoxide molecules were arranged in bi-stable positions and pushing one molecule would prompt a chain reaction of cascading motion like a domino.[9] Atomic-scale manipulation can be used to break or form chemical bonds and tailor chemical reactivity of the surface, based on which STM lithography was developed to pattern resist in atomic resolution. Simmons group demonstrated atomic scale devices on a hydrogenated Si(100) by selectively detaching hydrogen atoms and dosing phosphorous atoms on these sites.[11, 24] The work shed light on how atomic-scale manipulation can be applied to the fabrication of devices with designed quantum functions.

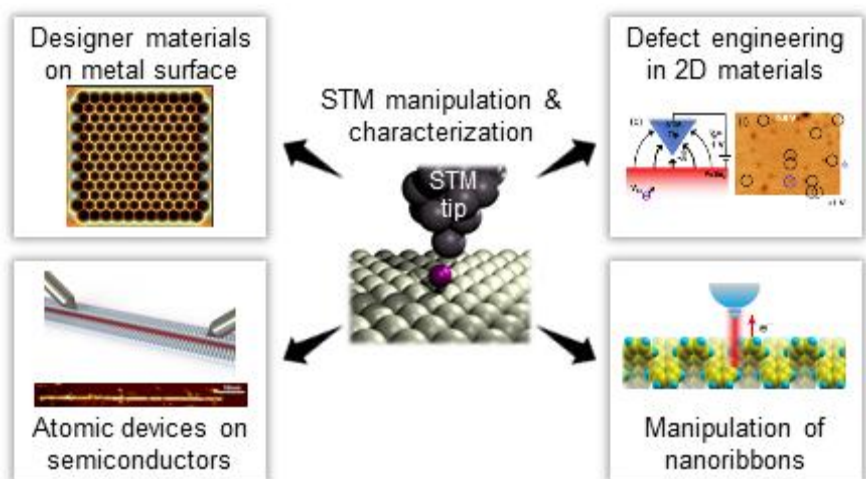

**Figure 2.** STM manipulation and *in situ* characterization applied to various materials systems. Clockwise from top-left: molecular graphene assembled on Cu(111) (adapted with permission.[21] Copyright 2012, Springer Nature); 3D manipulation of vacancies in PdSe$_2$ (adapted with permission.[25] Copyright 2018, American Physical Society); conversion of polymers to graphene nanoribbons by hole injection (adapted with permission.[26] Copyright 2017, Springer Nature, licensed under CC BY 4.0); multiprobe transport measurement on atomic wire patterned on hydrogenated Ge(100) (adapted with permission.[16] Copyright 2017, IOP publishing, licensed under CC BY 3.0)

For this rapidly developing field, a comprehensive review of all the progress is beyond the scope of this article. Here we review a few representative research accomplishments from us and our collaborators (**Figure 2**). First, STM manipulation was performed on noble metal surfaces with free-electron-like surface states. The manipulation of atoms and molecules adsorbed on these "mundane" materials can transform them into an artificial 2D material with distinct band structures. This method has shown as a perfect tool to design novel 2D materials with required electronic properties.[21, 27] Second, the electronic properties of real 2D materials, such as graphene and transition metal chalcogenides, were tailored by controlled manipulation of defects and dopants in them.[25, 28] Third, the bottom-up formation of graphene nanoribbons (GNRs) was controlled by STM to create precise heterostructures of polymer and GNR.[26, 29] Finally, STM manipulation was performed on semiconducting surfaces to realize atomic-scale functional systems, allowing implementing device functions and characterizing them *in situ*. A big obstacle for characterizing the device functions of these nanostructures are the fabrication processes that require exposure to the ambient environment, which can easily contaminate or destroy the atomically precise nanostructures and thus stymie the demonstration of the designed functions. The advent of multiprobe STM made it possible to both fabricate and characterize atomic structures *in situ* on semiconductor surfaces.[16] Continuous development of multiprobe STM allowed for detecting transport of charge and spin in the atomic scale,[30] and further application of these techniques will not only enable revealing of the transport of the quantum states but also provide a route to manipulate and control the structures to observe the responses of these states.

## 2. Experimental Method with Single- and Multi-probe STM

2.1. Principle of Atomic-Scale Manipulation with STM

The technique of atomic-scale manipulation is based on the fine tuning of the interactions between the substrate, adatoms, and STM tip. Various methods of the manipulation and the underlying principles have been discussed thoroughly in several articles and reviews already.[6, 31] Here we will give a brief summary and introduce some recent developments.

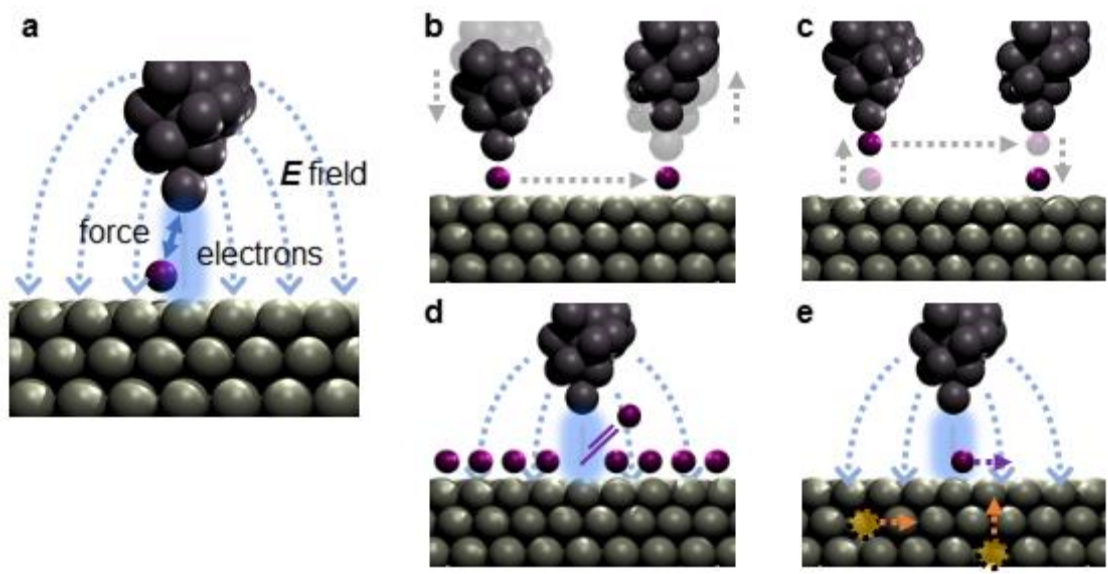

**Figure 3.** Tip-atom interaction and experimental schemes for atom manipulation. a) Physical mechanisms generating tip-atom interaction. Schematic of b) lateral manipulation, c) vertical manipulation, d) extraction of adatoms with bias pulses, and e) field-induced manipulation of adsorbates or subsurface defects.

Different mechanisms were proposed to explain interactions between the tip and atoms/molecules on the surface (**Figure 3**a). For examples, van der Waals and electric force can come into play between the tip and atoms, and tunneling electrons can transfer charge and energy from the tip to the atoms. Typically, the manipulation is a result of the participation of several of these interactions. Even a simple lateral manipulation of an atom on a metal surface involves several mechanisms depending on the tip-sample distance, bias voltage, and tunneling current.[31d, 31e] Figure 3b illustrates a typical method of lateral manipulations, where an STM tip pushes or pulls the adsorbates on the surface. Importantly, lateral manipulation uses gentle bias less than tens of millivolt not to excite the adsorbates and keep the atoms adsorbed on the surface.[2] However, vertical manipulation in Figure 3c intentionally excites the adsorbates with a large bias voltage of a few volts to transfer adsorbates between the tip and the sample. A large bias voltage pulse can break the bond of adsorbates with the sample and attach them to the tip. The adsorbates stuck to the tip are moved to a new position and then dropped by another bias voltage.[17] Large bias voltage on the STM tip generates an electric field or a tunneling current that can dissociate chemical bonds between the adsorbates and the surface, and cause their diffusion or complete desorption (Figure 3d).[5, 32] The method was applied to pattern hydrogenated semiconductor surfaces by detaching hydrogen atoms, where the efficiency and the precision of the patterning process show a strong dependence on bias voltage and tunneling current.[31f]

Elucidation of the manipulation mechanisms and the discovery of novel materials offer opportunities to explore new schemes of atom manipulation. One example is the demonstration of 3D manipulation of defects in 2D layered materials (Figure 3e). For example, STM was used to visualize and manipulate charge states of defects in the bulk insulating hexagonal boron nitride (hBN) film covered by single-layer graphene on the tip.[28b, 33] Graphene is conductive enough for STM imaging, but do not completely screen the electric field from the tip penetrating into the hBN. The manipulated defects modified the local doping levels of graphene, which created p-n junction or quantum dots.[28a, 34] Furthermore, 3D manipulation of individual vacancies was demonstrated in $PdSe_2$, a newly discovered pentagonal layered material.[35] Here, the electric field from the tip changed the charge states of individual vacancies in the bulk and enabled to write and erase vacancies at particular lattice positions.[25] These examples demonstrate how new materials and new manipulation technique are combined to reveal new physical processes and create new quantum states.

## 2.2. *In situ* Characterization of the Electronic Structure

Scanning tunneling spectroscopy (STS) measures the differential conductance d$I$/d$V$, and maps the electronic local density of states (LDOS) with high spatial and energy resolutions (**Figure 4**). STS, when combined with atomic scale manipulations, enables STM to image, manipulate, and characterize the atomic structures in the same experimental platform. For example, Figure 4b shows a circular quantum corral fabricated on a Cu(111) surface and its d$I$/d$V$ spectrum, where strong peaks appear at certain energies due to the quantum confinement of the surface states.[4, 36] In principle, the shape of corrals can be varied to obtain a desirable energy spectrum by design.

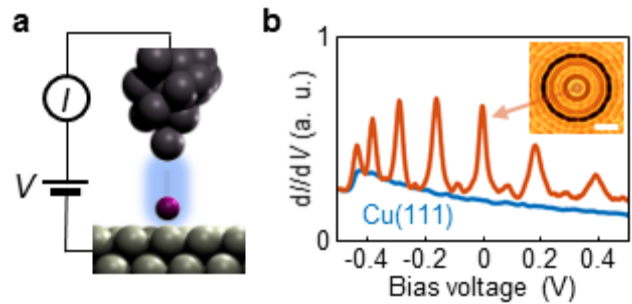

**Figure 4.** Electronic characterization with single-probe STM. a) Schematic of STS measurement. b) d$I$/d$V$ spectra taken over clean Cu(111) (blue line) and the center of a quantum corral (orange line). Inset shows a topographic iamge of the quantum corral ($V$ = 10 mV, $I$ = 1 nA, scale bar corresponds to 5 nm).

Integration of multiple probes in the same STM platform greatly enhances the capability of characterization of electronic properties (**Figure 5**). The state-of-the-art multiprobe STM has four STM scanners that can be operated independently with atomic resolution. These probes can characterize samples in multiple modes, mapping electronic density of states in tunneling mode, serving as source and drain electrodes in contact mode or floating electrodes for probing field effect transconductance.[30d, 37] Each tip can be navigated with atomic precision to allow precise contact on the sample. Moreover, atomic scale manipulations can be performed with the multiprobe STM and then characterized *in situ*.[16]

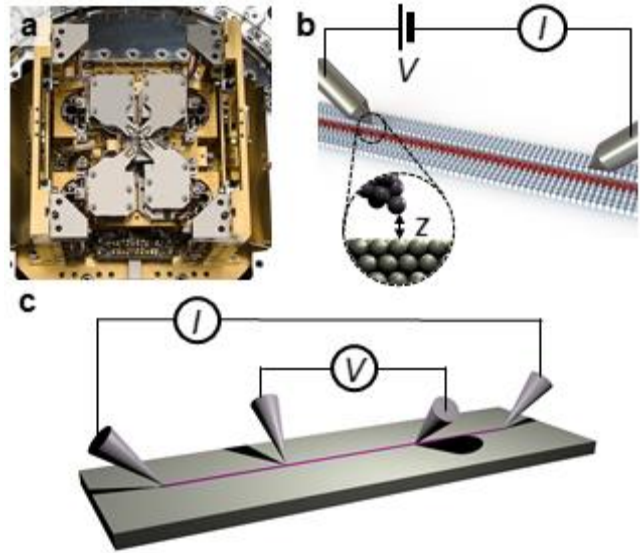

**Figure 5.** Characterization of transport properties with multiprobe STM. a) A photo of multiprobe STM scanners (image courtesy of Scienta Omicron GmbH, Germany). b) Two-probe transport measurement scheme (adapted with permission.[16] Copyright 2017, IOP publishing, licensed under CC BY 3.0). c) Four-probe transport measurement scheme.

Specifically, multiprobe STM can perform two-probe and four-probe transport measurement on the structure made by atomic-scale manipulation protocols.[30d] Two-probe method (Figure 5b) is the simplest way to measure the transport, but the measured *I-V* contains contributions from the sample-tip contact resistance. To remove the effect of the contact resistance, the four-probe method is used, where two outer probes supply the current while two inner probes measure the voltage (Figure 5c). Although the four-probe transport measurement is an ideal method to characterize conductance of the mesoscopic systems, it is hard to apply the method to atomic scale structures because of the relatively large tip radius (tens of nanometers). One possible solution to overcome such limitation is scanning tunneling potentiometry.[30d] The method utilizes two probes to supply a current and a third probe to scan in between that measures the local potential. It fully takes advantage of the atomic resolution of STM and successfully visualized electron scattering around atomic defects.[38] However, from a fundamental point of view, the determination of transport properties at the atomic scale ultimately requires understanding of contact formation between the tip and the sample in atomic precision, and thus new methodologies are necessary. One method is adopting STS into the two-probe configuration for two-probe STS (2P-STS), where both STM tips are kept in tunneling or single-atomic contact regime during the two-probe transport measurement.[16, 39] The state-of-the-art multiprobe STM can bring two probes at the separation of a few tens of nanometers, and keep the tip position and height constant with picometer precision so that the tunneling junctions have well-defined *I-V* characteristics to remove the uncertainty of contact resistance in conventional two-probe measurement (Inset of Figure 5b). Combined with atomic scale manipulations, these multiprobe methods make it possible to detect and control the transport of quantum states.

## 3. Creating Designer Quantum States on the Metal Surfaces

Single atoms and molecules deposited on the surfaces of metal substrates can be easily imaged with STM. Some atoms and molecules are adsorbed weakly enough for STM manipulation. The first demonstration of the atom manipulation with an STM tip was performed with Xe atoms adsorbed on a Ni(110) surface.[2]

However, the large number of bulk carriers in metals can suppress the electronic properties of the atomic structures fabricated by atom manipulation. In this regard, it is desirable to use noble metals that have a bulk band gap at the $\Gamma$ point of Brillouin zone, and Shockley surface states which have free-electron-like band around the Fermi level.[40] In this case, the surface states behave essentially like a two-dimensional electron gas (2DEG) and are well-decoupled from the bulk states with a long coherence length that can approach hundreds of nanometers.[41] Surface electrons scattering off defects can create standing wave-like interference patterns, which can be visualized by STM.[42] Atoms and molecules adsorbed on the surface also behave as scattering centers for the surface states. Using the atom manipulation method, artificial atomic structures have been fabricated to guide surface electrons to create new electronic states. Quantum corrals, which are cages made out of the atoms or molecules, showed vividly the confinement of electron wavefunction and the formation of discrete energy levels of the surface electrons.[4] More complicated atomic structures were constructed to further manipulate quantum states, such as a mirage of quantum states in elliptical quantum corrals,[43] extraction of the phase of electron wavefunction from isospectral quantum corrals,[44] and holographic encoding of information in the energy dimension in addition to the spatial dimensions [45]. These seminal works have demonstrated the capability of atom manipulations in confining surface electrons to give rise to interesting quantum phenomena previously only being conceived by theory.

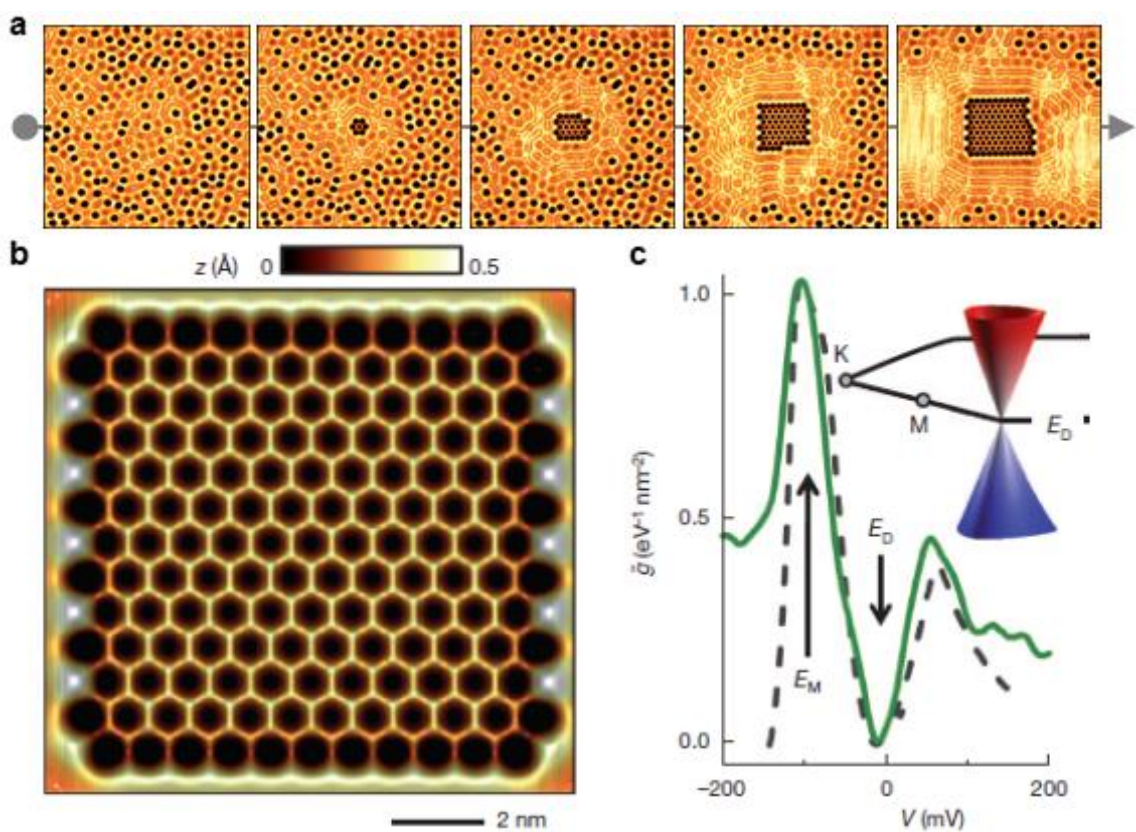

**Figure 6.** Molecular graphene made of CO molecules assembled on Cu(111). a) Sequence of STM topographic images for assembling CO molecules into molecular graphene. b) Topographic image of the final assembly of molecular graphene. c) Normalized conductance

spectrum taken at the center of the molecular graphene (solid line). The dashed line shows LDOS calculated by the tight-binding model. The inset on the right top shows the band structure from the tight-binding model. Adapted with permission.[21] Copyright 2012, Springer Nature.

Recently, the atom manipulation on noble metal surfaces allowed to realize designer materials with artificial lattice structure and distinct electronic structure. A prominent example is molecular graphene built by Gomes et al., which is a graphene-like structure fabricated by atom manipulation of CO molecules on Cu(111) (**Figure 6**).[21] CO molecules are relatively easy to manipulate, and their interaction with the surface states is well-understood where CO molecules behave as repulsive potential for the surface states [46]. Figure 6a shows the sequence of assembly of CO molecules to the final structure of graphene in Figure 6b. When CO molecules were arranged in a triangular lattice, surface electrons were confined between the molecules to form a honeycomb lattice. Assembled structures transformed the electronic band from a free-electron-like 2DEG to a graphene-like band with massless Dirac fermions. Differential conductance spectrum taken in the middle of the lattice showed a V-shape dip at $E_D$ originating from the Dirac cone in the band (Figure 6c). A peak at $E_M$ appeared right next to the Dirac point $E_D$ arising from the van Hove singularity at M point in Brillouin zone. The work demonstrated a proof-of-principle concept that artificial lattice structure can be assembled by atom manipulation to realize designer materials. The designer materials give a wide tunability of the interaction between the atoms and molecules. In molecular graphene, it was demonstrated that doping, effective mass, and strain could all be tuned to create novel topological phases such as massive Dirac fermions from Kekulé distortion and even strain-induced quantum Hall states.[21, 47]

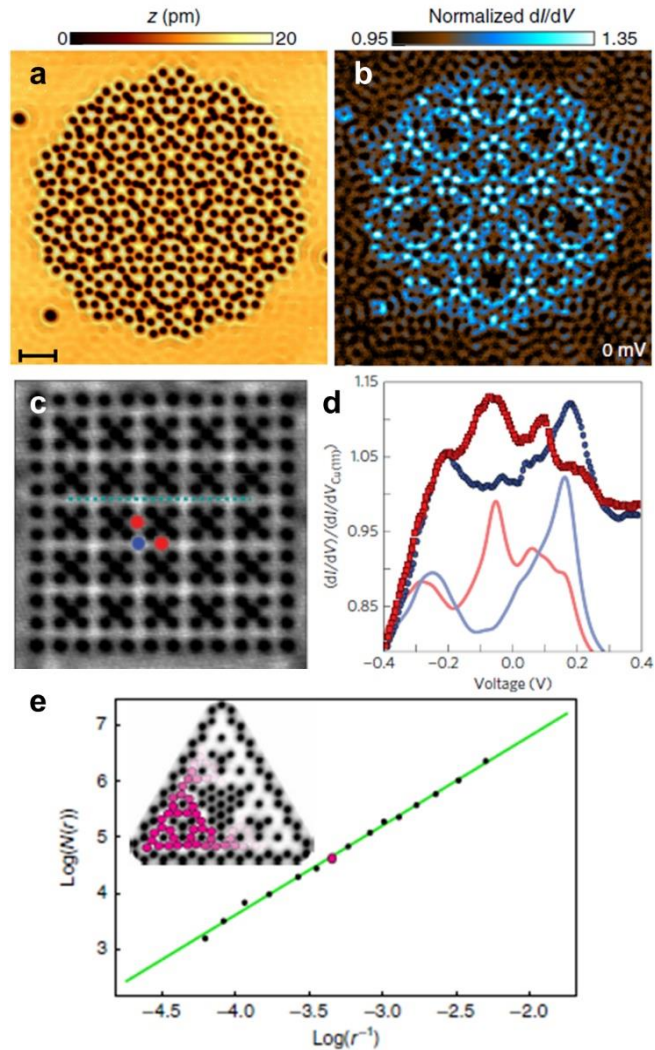

**Figure 7.** Various designer materials built by CO molecules on Cu(111). a) Topographic image of Penrose tiling quasicrystal, and b) d$I$/d$V$ map obtained at the same location with $V$ = 0 mV. Panels (a,b) are reprinted with permission.[27a] Copyright 2017, Springer Nature, licensed under CC BY 4.0. c) Topographic image of Lieb lattice. d) d$I$/d$V$ spectra taken at two non-equivalent sites in Lieb lattice (dotted lines), and theoretical fitting with tight-binding calculation (solid lines). Panels (c,d) are reprinted with permission.[27b] Copyright 2017, Springer Nature. e) Topographic image of Sierpinski triangle fractal (inset) and box fitting for determining the dimension of electron wavefunction (reprinted with permission.[27c] Copyright 2018, Springer Nature).

The designer materials further allowed the demonstration of hypothetical materials that do not usually exist in nature. **Figure 7** shows three examples of such hypothetical materials realized by CO molecules assembled on Cu(111). The first example is a quasicrystal from Penrose tiling realized by Collins et al. (Figure 7a).[27a] Penrose tiling is a popular aperiodic tiling that can fill the surface completely without translational symmetry. By positioning CO molecules in the Penrose tiling pattern, a quasicrystal with five-fold symmetry was constructed. Because the band theory description of electronic structure is based on the translational symmetry, it cannot describe the electronic structure of quasicrystal. Instead, the authors classified the atomic sites with their vertex structures relative to the neighboring sites. By measuring d$I$/d$V$ maps, they showed that the electronic structure strongly depends on the first order vertex structure (Figure 7b). The second example if the realization of Lieb lattice by Slot et al., based on a theoretical model proposed to explain magnetism from the formation of flat band (Figure 7c).[27b] The lattice has two distinct atomic sites, one in the corner (blue dot in Figure 7c) and the other at the edge (red dot in Figure 7c). Especially, wavefunctions of the flat band were shown to locate at the edge sites, and d$I$/d$V$ spectra clearly showed peaks only for the edge sites. The experiment matched nicely with the tight-binding calculations (Figure 7d). The third example is the Sierpinski triangle fractal realized with CO molecules on Cu(111) by Kempkes et al.[27c] Fractals have self-similar structure with non-integer dimensionalities, and the Sierpinski triangle is supposed to have a dimension of log(3)/log(2) ≈ 1.585. It was unknown, however, whether the electrons inside the Sierpinsky triangle would also have the same non-integer dimensionality. The authors built the lattice with CO molecules up to the third generation, and measured d$I$/d$V$ spectra that showed the electron wavefunctions with a dimension of about 1.58 in the measured energy range of -0.3 ~ 0.3 V (Figure 7e).

The structures of these examples were all first proposed as a theoretical model. It was the atom manipulation that made it possible to create the proposed structures and observe the electronic properties. By fabricating designer materials, these examples demonstrated the transformation of the use of STM from a characterization tool to a promising manipulation platform to realize designer quantum states of condensed matter systems.[48]

## 4. Controlling 2D materials by Manipulating Atomic Defects

The advent of 2D materials opened exciting new opportunities for defect engineering. Because of the low dimensionality of the materials, atomic defects such as vacancies and dopants can affect the electronic structures more dramatically than in 3D materials.[49] STM is a great tool for defect engineering as it not only allows imaging defects at the atomic scale but also allows manipulating them individually.

The graphene surface offers an ideal platform for atom manipulation that can be combined with tunable doping effects from a back-gate electrode. Using an STM tip, Wang et al. managed to assemble Co and Ca adatoms on graphene.[28c, 28d] While Co adatoms were manipulated by dragging them with conventional lateral manipulation method (Figure 3b),[28c] Ca impurities were only successfully manipulated by pushing with bias pulses on the STM tip combined with a proper back-gate voltage (Figure 3e and **Figure 8**a-c).[28d, 50] The Ca dimers were positively charged and the high electric field from the bias pulses generated a repulsive force to Ca dimers. However, they were only moved when a negative back-gate voltage was applied to tune graphene to a heavily p-doped state. The pushing of Ca dimers with the STM tip did not work in n-doped graphene. It was assumed that highly p-doped graphene reduced the interaction with positively charged Ca dimers, while n-doped graphene attracted Ca dimers and made them too sticky to move.

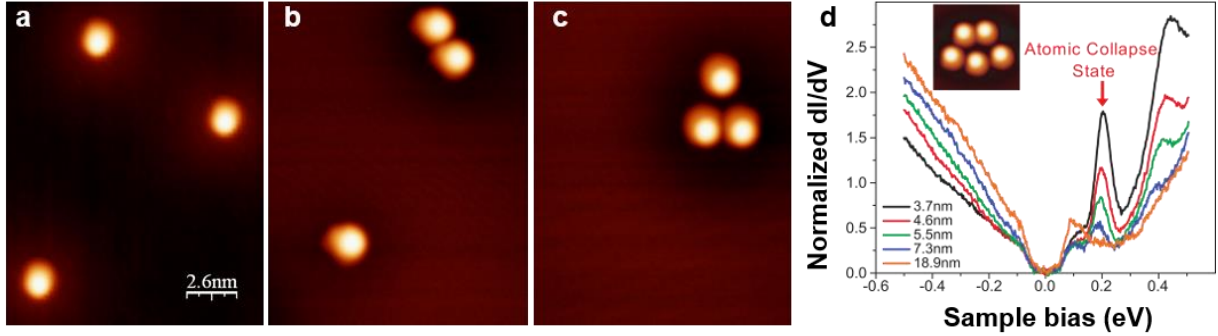

**Figure 8.** a-c) STM topographic images presenting the process of manipulating Ca dimers. The tunneling parameters were first set to $V_s = 0.5$ V, $I_t = 2$ pA and $V_g = -60$ V. The tip was placed near the dimer on the side opposite to the desired direction of motion. The Ca dimer was pushed forward by applying a negative sample bias pulse of $V_s = -1$ V (adapted with permission.[50] Copyright 2017, Dillon Wong). d) d$I$/d$V$ spectra at different distances from the center of a five Ca dimer cluster. Atomic collapse states are marked by a red arrow (reprinted with permission.[28d] Copyright 2013, AAAS).

Through the atom manipulation, the interaction of ultra-relativistic Dirac fermion with the tunable charged impurities was studied. It was revealed that electrons and holes responded differently to the Coulomb potential. The observed asymmetry was used to explain the differences in the mobility between electrons and holes in transport measurements.[28c] Furthermore, Ca dimer clusters acted as artificial nuclei whose charge could surpass the supercritical limit, allowing direct observation of the transition from a subcritical behavior to a supercritical atomic collapse behavior when nuclear charge increased (Figure 8d).[28d] Later, Mao et al. demonstrated that the supercritical charge could also be built up in single-atom vacancy in graphene by applying tip pulses.[51]

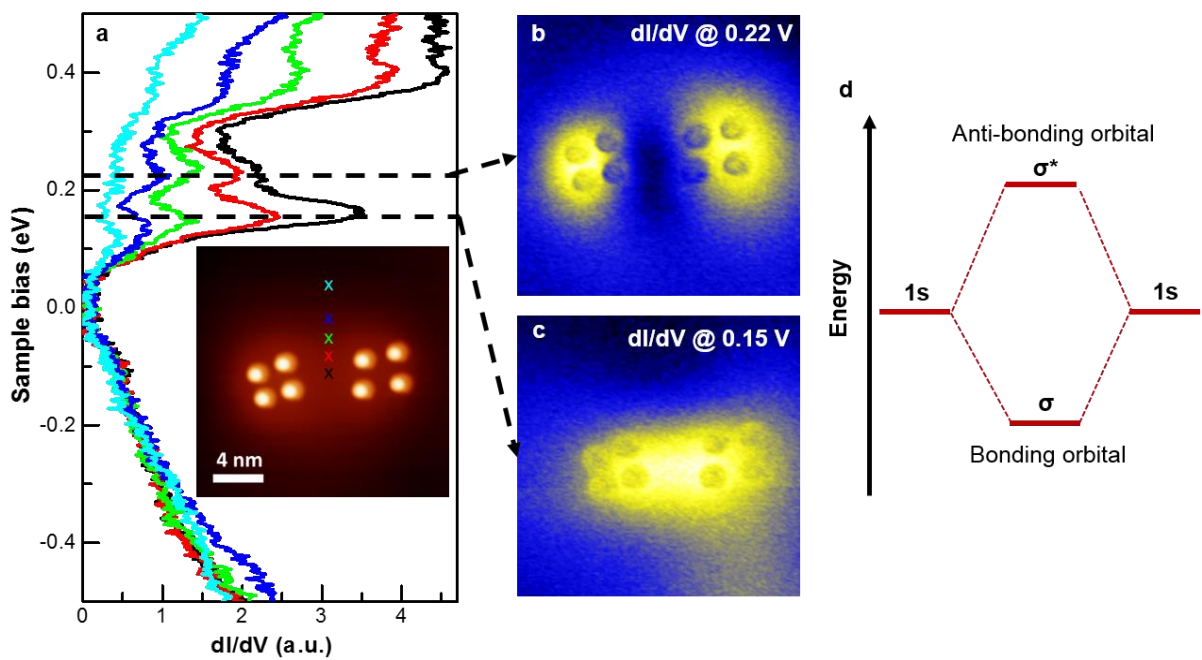

**Figure 9.** a) d$I$/d$V$ spectra acquired in between two clusters composed of four Ca dimers (STM image in the inset). b,c) d$I$/d$V$ maps at the sample biases of 0.22 V and 0.15 V, respectively. Panels (a-c) are adapted with permission.[50] Copyright 2017, Dillon Wong. d) Schematic bonding orbitals of the atomic collapse molecule.

In addition, atom manipulation was used to build two clusters of Ca dimers with atomic collapse orbitals close to each other (inset in **Figure 9**a). The d$I$/d$V$ spectra taken in between two clusters showed two resonant peaks at 0.15 V and 0.22 V. The d$I$/d$V$ maps at 0.15 V and 0.22 V suggested that they corresponded to bonding and anti-bonding orbitals, respectively (Figure 9b,c). This result implies that two close artificial atoms of Ca clusters can form a bond to become an atomic collapse molecule. The bonding orbitals are similar to the case of a hydrogen molecule as illustrated in Figure 9d. Therefore, the atom manipulation in graphene opens a new possibility for engineering new artificial atoms/molecules to realize exotic quasi-bound states. The array of artificial atoms would guide the motion of the carriers in graphene, which may be applied to future graphene-based devices.

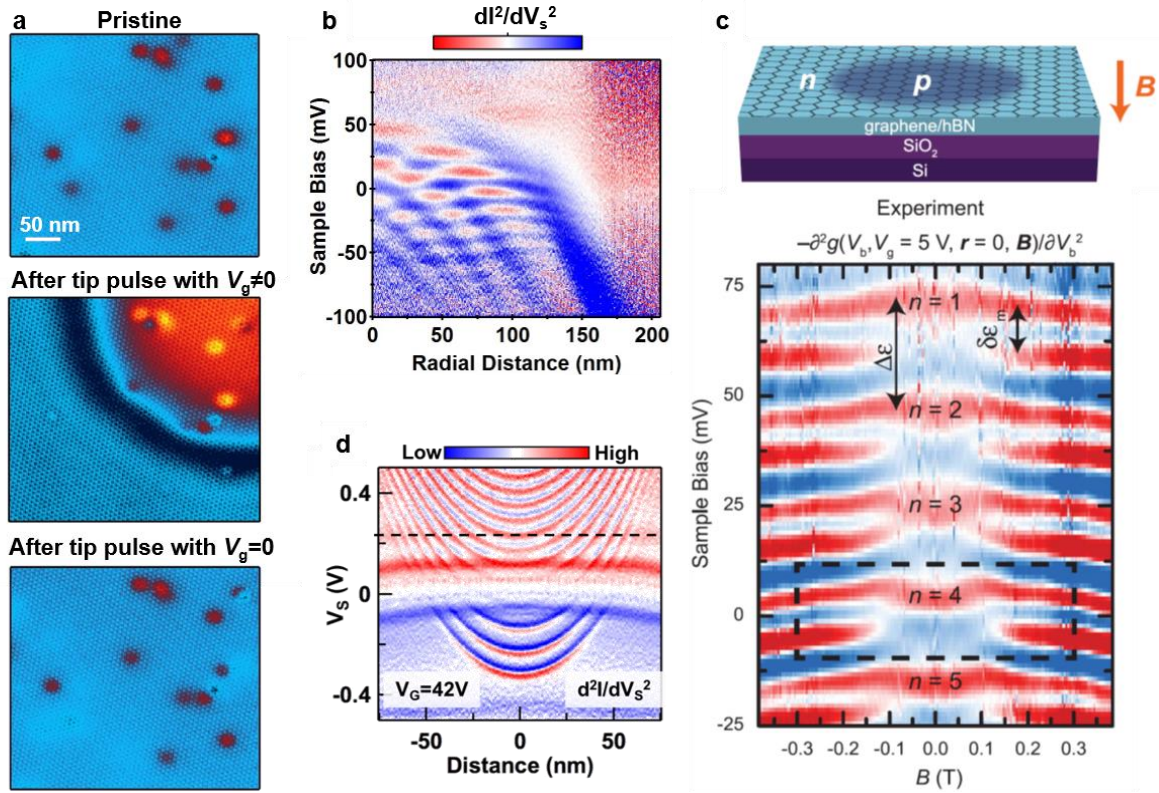

**Figure 10.** a) Nanoscale doping patterns controlled by bias pulses on the STM tip (reprinted with permission.[28b] Copyright 2016, American Chemical Society). b) Spatially resolved energy levels inside a circular graphene quantum dot (reprinted with permission.[34] Copyright 2016, Springer Nature). c) Differential tunneling conductance map versus magnetic field which reveals the Berry phase switching at a magnetic field of about 0.11 T (reprinted with permission.[28a] Copyright 2017, AAAS). d) Spatially resolved energy levels of a circular bilayer graphene quantum dot (reprinted with permission.[52] Copyright 2018, American Chemical Society).

The atomic scale manipulation with bias pulses on the STM tip has been extended to another spatial dimension perpendicular to the sample surface so that 3D manipulation is possible. In the heterostructure of single-layer graphene atop bulk hexagonal boron nitride (hBN), charge states of the defects in hBN were toggled by injecting or extracting trapped charges by the electric field from the STM tip.[33] Using the back-gate voltage during the bias pulse, the charge states of hBN defects could be controlled which allowed tuning of the local graphene doping effect.[28b] The technique provided a reversible control of doping patterns in graphene at the nanometer scale, which was utilized to create a stationary circular p-n junction and quantum dot (**Figure 10**a). STS allowed mapping the electronic structure of tunable graphene quantum dots (Figure 10b).[34] External magnetic field revealed the Berry phase associated with the topological properties of Dirac fermions in graphene (Figure 10c).[28a] The method was also applied to bilayer graphene to study the behavior of massive Dirac fermions confined within circular p-n junction (Figure 10d).[52] This demonstrated, in principle, that the technique can be employed on different kinds of 2D material heterostructures.

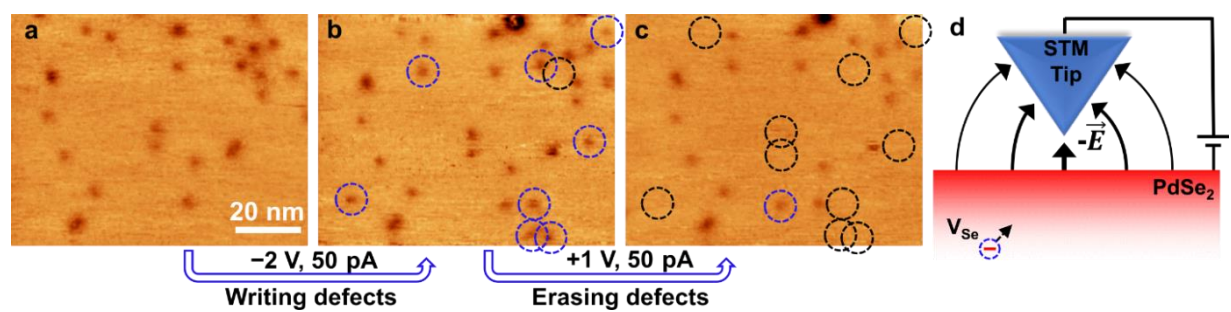

**Figure 11.** a-c) STM topographic images presenting the "writing" and "erasing" process of Se vacancies on the PdSe$_2$ surface. Blue dashed circles mark the new vacancies after the writing process; black circles indicate disappeared vacancies after erasing process. d) Schematic illustration of the "writing" process. A negatively charged vacancy below the surface migrates toward the STM tip by the electrical field. Reprinted with permission.[25] Copyright 2018, American Physical Society.

Indeed, Nguyen et al. showed the 3D manipulation of individual Se vacancies in PdSe$_2$.[25] An STM tip was used to "write" and "erase" of near-surface vacancies through controlling Se vacancy migrations vertically and laterally (**Figure 11**). $PdSe_2$ was chosen because of its unique puckered pentagonal structure with low lattice symmetry.[35] In contrast to typical hexagonal TMDs, the distinct pentagonal network in $PdSe_2$ makes covalent bond strength relatively weak, which facilitates the migration of Se vacancies. In the manipulation of Se vacancies, an STM tip act as a nanoscale movable electrostatic gate. The "writing" process was performed by scanning over a surface area at high negative sample bias. Consequently, new Se vacancies were created under the scanning area (Figure 11b). To "erase" these defects, the STM tip scanned over the area again but with an opposite positive sample bias (Figure 11c). The procedure had very high efficiency, where almost 90 percent of the newly created Se vacancies in the "writing" process were removed after the "erasing" treatment.

The underlying physical mechanism of the reversible "writing" and "erasing" processes are believed to originate from an effect of both the electric field and tunneling electrons. At the negative sample bias, Se vacancies become negatively charged by tip-induced band bending.[25] During the scanning at a larger negative sample bias, the electric field between the tip and the sample drives negatively charged Se vacancies towards the tip as illustrated in Figure 11d. In an "erasing" process, the sample bias was switched to a positive bias of +1 V. At this positive bias, the Se vacancies became neutral, so the role of the electric field can be neglected. Hot electrons tunneling from the tip can play a role in this case, where an inelastic scattering process may trigger the vacancies to migrate deeper into the bulk. Because the vacancies tend to diffuse to the part with lower formation energy, the reduction of the density of near-surface Se vacancies after scanning at the positive bias indicates lower formation energy of Se vacancies in subsurface compared to the surface. This result shows the possibility of performing 3D manipulation of vacancies in a few-layer $PdSe_2$ to create a reversible doping pattern at the nanoscale.

## 5. Enabling Chemical Reactions from Organic Molecular Precursors

As discussed above, STM has been widely used in characterizing and manipulating on-surface molecules. The inelastic electron tunneling (IET) with STM was an important breakthrough for molecule manipulation (**Figure 12**a), because IET through an adsorbate induces vibrational excitations that can cause atom/molecule desorption,[53] structural change[54] or dissociation.[32, 55] It became a powerful tool for "molecular surgery" and inter-molecular reactions since the 1990s. More importantly, the corresponding reaction pathways can be determined by imaging the intermediates and final states in the tip-induced reaction processes. In 2000, Laohon et al. reported a multistep unimolecular reaction by breaking the C-H bonds in HCCH molecules (Figure 12b). Combining it with the IET spectroscopy (IETS), they determined the vibrational modes and bonding geometry accompanied with HCCH dissociation.[56] At the same year, Hla et al. showed the dehalogenation and lateral manipulation of single molecules with STM, achieving Ullmann coupling between two iodobenzene molecules adsorbed at a Cu(111) step edge (Figure 12c).[57] In 2005, by breaking two C-H bonds at once, CoPc molecules were sequentially dehydrogenated where Kondo effect was controllably realized in the dehydrogenated CoPc (d-CoPc) (Figure 12d).[58] More recently, via the similar STM tip-assisted dehydrogenation or dehalogenation, arynes,[59] triangulene[60] and a reversible Bergman cyclization action (Figure 12e)[61] were controllably generated and characterized. On a large bandgap semiconductor of the rutile $TiO_2(110)$ surface, Tan et al. studied the tip-induced reactions of various molecules, such as $O_2$[62] and $CO_2$,[63] and $H_2O$.[64] Recently, they revealed a new mechanism for bond-selective reactions in single $CH_3OH$ molecules, that the breaking of C-O, C-H, and O-H bonds can be accurately controlled by selective injection of electrons and holes from a tip as shown in Figure 12f.[65]

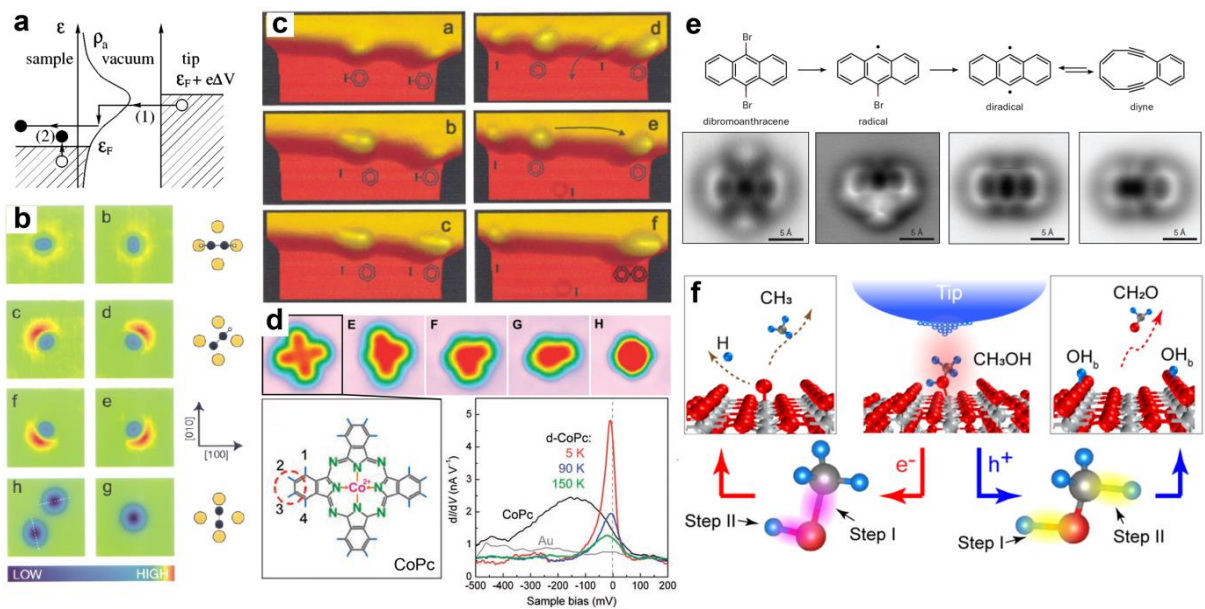

**Figure 12**. STM tip-assisted manipulation of various organic molecules. a) Schematic of inelastic electron tunneling. Electrons tunneling to an adsorbate-induced resonance with density of states $\rho_a$ induce vibrational excitations, while electronic excitations within the substrate induce vibrational relaxations (reprinted with permission.[32] Copyright 1997, American Physical Society). b) STM images of DCCD, CCD, and CC molecules during the tip-assisted manipulation (reprinted with permission.[56] Copyright 2000, American Physical Society). c) STM images showing the initial steps of the tip-induced Ullmann synthesis between two iodobenzene molecules adsorbed at a Cu(111) step edge (reprinted with permission.[7] Copyright 2000, American Physical Society). d) STM images of a single CoPc molecule during the dehydrogenation process from an intact CoPc to d-CoPc. Hydrogen atoms 2 and 3 in one lobe were dissociated in the experiments (see structural formula at bottom-left). d$I$/d$V$ spectra showed Kondo resonance from d-CoPc at different temperatures (reprinted with permission.[58] Copyright 2005, AAAS). e) A reversible Bergman cyclisation action realized during STM tip-assisted dehalogenation from a 9,10-dibromoanthracene molecule (reprinted with permission.[61] Copyright 2016, Springer Nature). f) Schematic of the distinct reaction processes after the injection of electrons or holes during the tip-induced

dissociation of $CH_3OH$ on a rutile $TiO_2(110)$ surface (reprinted with permission.[65] Copyright 2018, American Chemical Society).

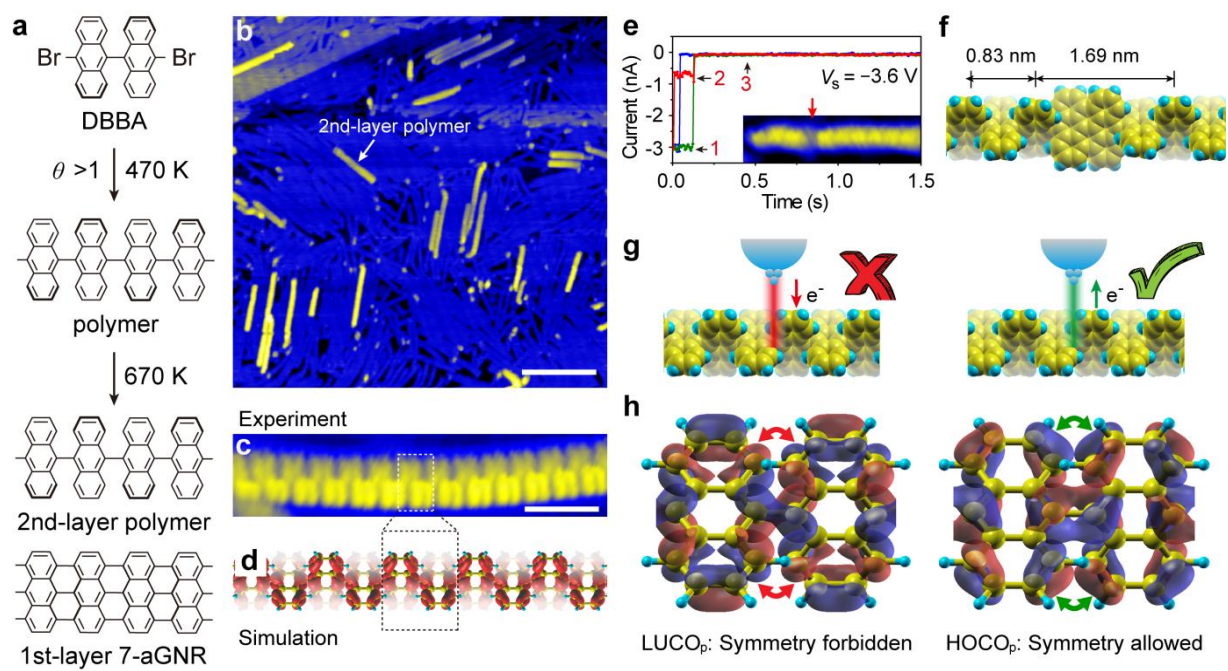

**Figure 13**. STM tip-assisted controllable conversion of quasi-freestanding polymers to GNRs. a) Sketch of the synthesis of the 2nd-layer polyanthrylene chains on 7-aGNRs from DBBA molecules with stepwise annealing at 470 and 670 K. b) A large-area STM image showing the polymer chains on 7-aGNRs. c) A high-resolution STM image showing the detailed structure of the polymer ($V_s = -2$ V, $I_t = 10$ pA, scale bar is 2 nm). d) Charge density distribution of the $HOCO_p$. Dashed box marks the single unit in the polymer. e) Three typical tunneling current $vs$. time ($I_t$-$t$) curves during the bias pulse of $V_s = -3.6$ V with feedback loop off (set point $V_s = -2$ V, $I_t = 100$ pA). Inset is an STM image of GNR segment (red arrow) in a polymer chain formed by the bias pulse. f) Atomic structure of a polymer chain with a short GNR segment. g) Schematic of injecting electrons (left) and holes (right) from an STM tip to the polymer site. Local conversion can be achieved with hole injection but not with electron injection. h) Simulated charge density distributions of $LUCO_p$ and $HOCO_p$ in the polymer where blue and red colors indicate the different sign of wavefunctions. The out-of-phase overlap (i.e.,

opposite sign) in LUCO$_p$ is marked by red arrows and the in-phase overlap (i.e., same sign) in HOCO$_p$ is marked by green arrows, indicating that the C-C bond formation in the cyclodehydrogenation is symmetry forbidden in LUCO$_p$ but symmetry allowed in HOCO$_p$. Adapted with permission.[26] Copyright 2017, Springer Nature.

By injecting charge carriers with an STM tip, Ma et al. recently demonstrated *in situ* manipulation and characterization of polymer-to-graphene nanoribbon (GNR) reactions, which gave various intraribbon heterojunctions (HJs) with controllable interfaces for designer functionalities.[26, 29a] The armchair GNRs with a width of seven carbon (7-aGNRs) were usually synthesized through a two-step annealing process from 10,10'-dibromo-9,9'-bianthryl (DBBA) molecules.[66] As illustrated in **Figure 13**a, when the effective coverage of molecular precursors is larger than 1 ML, the molecules on the second-layer (2$^{nd}$-layer) can polymerize to form polyanthrylene chains but can not cyclodehydrogenate even when annealing at 670 K. (Figure 13b). The results confirmed that the catalytic effect of the metal surface was critical for the conversion of the polymer to the GNR by cyclodehydrogenation. The high-resolution STM image of a 2$^{nd}$-layer polymer chain, shown in Figure 13c, is consistent with the simulated features in Figure 13d. Interestingly, the STM tip can trigger the cyclodehydrogenation in a controllable manner at an arbitrary site in the 2nd-layer polymer chain. As presented in Figure 13e, after a pulse treatment at a bias of −3.6 V, the red arrow marks the site that was converted to GNR in the polymer with the corresponding structural model shown in Figure 13f. The representative $I_t$-$t$ curves showed three typical current terraces, suggesting a three-step reaction process during the pulse treatment. Based on this result, the corresponding reaction pathway was proposed.[26]

Bias dependence of the polymer-to-GNR reaction has given the insight into the manipulation mechanism. Only the negative bias (hole injection) was found to trigger the polymer-to-GNR reactions, while the positive bias (electron injection) did not trigger this reaction and only broke the polymer chains, as illustrated in Figure 13g. Based on the nudged elastic band (NEB) calculations, the injection of two holes significantly reduced the reaction barrier, while two electrons did not change the reaction barrier. This is consistent with the Woodward-Hoffmann rules for orbital symmetry conservation in pericyclic reactions.[67] As shown in Figure 13h, the formation of a C–C bond through electron injections into the lowest unoccupied crystal orbital of the polymer ($LUCO_p$) state is symmetry forbidden (red arrows, Figure 13h) due to the opposite phases of wavefunctions, while it is symmetry allowed (green arrows, Figure 13h) through hole injections into the highest occupied crystal orbital of the polymer ($HOCO_p$) state as the involved wavefunctions have the same phases. This hole-assisted cyclodehydrogenation is similar to the well-known Scholl reaction.[68] In organic chemistry, oxidants such as $FeCl_3$ are often used to extract electrons (inject holes) in the Scholl reaction,[69] with which GNRs have been synthesized in liquid.[70] The ability to control the cyclodehydrogenations at selected molecular sites with an STM tip, even without a catalytic metal substrate or oxidants, provides an opportunity to synthesize freestanding GNRs and create novel intraribbon heterojunctions bottom-up.

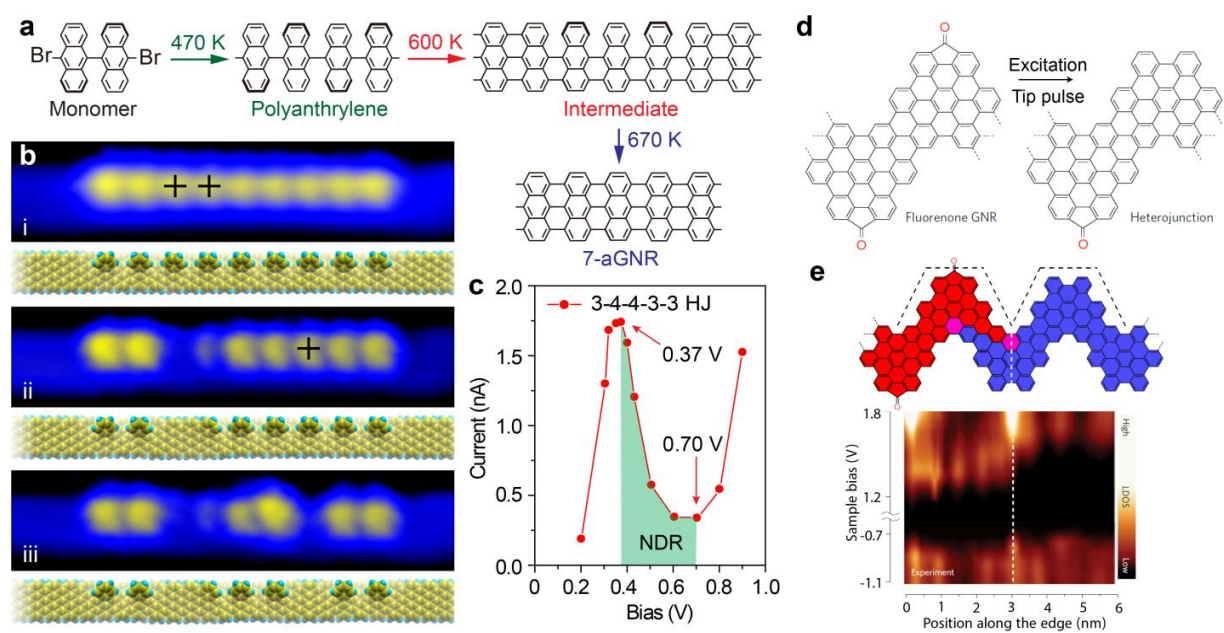

**Figure 14**. STM tip-assisted fabrication of functional GNR HJs. a) Temperature-dependent growth process of bottom-up synthesis of the 7-aGNRs using DBBA molecules as precursors, with an intermediate state with one-side cyclodehydrogenation formed at 600 K. b) Sequential direct writing of GNR segments in the intermediate to form HJs at designed molecular sites marked by crosses. Panels (a,b) are adapted with permission.[29a] Copyright 2018, American Physical Society. c) Calculated *I-V* curve of the device composed of graphene electrodes and the experimentally achieved 3-4-4-3-3 HJ in between. The numbers indicate the length of GNRs or intermediate segments in the anthrylene unit. The region showing NDR is marked with green shading (reprinted with permission.[71] Copyright 2018, Wiley). d) Schematic representation of the fabrication of a fluorenone/unfunctionalized chevron GNR HJ from a uniform fluorenone GNR. e) Determination of the band alignment of the p-n junction based on d*I*/d*V* spectra recorded along the edge of a heterojunction interface (black dashed line in the structural model). Panels (d,e) are reprinted with permission.[29b] Copyright 2017, Springer Nature.

Although the work on polyanthrylene chain demonstrated the proof-of-principle of the conversion of polymer-to-GNR by STM manipulation, the manipulation method requires further improvement because the high bias can easily break the polymer backbone before becoming the GNR.[26] Indeed, utilizing intermediate segments, where one side of the polyanthrylene is converted to the GNR while the other side remains in the polymeric structure, the treatment can greatly improve its reliability. As shown by Ma et al., the preexisting C-C backbone structure on the graphitized side of the intermediate led to a better tolerance of the bias treatment, and thus allowed a better control in creating designer GNR/intermediate HJs.[29a] To form the intermediate chains, a lower temperature of 600 K was used for the graphitization process (**Figure 14**a). After a pulse with a negative bias applied from an STM tip to the polymeric side of the intermediate segment, a fully graphitized GNR segment was created in the intermediate.[26, 72] By repeating this approach, polymeric units in the intermediate can be selectively converted to GNRs, generating complex GNR/intermediate HJs. An example is shown in Figure 14b, where a five-segment HJ is achieved. Because the 7-aGNR and intermediate segments have different band gaps, electronic properties of HJs can be modified by designing graphene and intermediate segments to achieve specific device functions. For example, the 5-segment HJ shown in Figure 14b is expected to display strong negative differential resistance (NDR) in *I-V* characteristic based on non-equilibrium Green's function calculation (Figure 14c).[71] A pronounced NDR appears at a relatively small bias with a large peak-to-valley current ratio (PVCR). Similar STM manipulation concept can be employed to fabricate other atomic-scale devices from molecular precursors.

Nguyen et al. reported another example of intraribbon HJs formed in chevron-like GNRs with STM-based manipulation (Figure 14d-e).[29b] The atomically precise GNR HJs were fabricated from a single precursor using two post-growth modification process, thermal annealing and tip pulses, that caused the cleaving of the sacrificial groups in GNRs for forming HJs. Because STM manipulation can create precise GNR HJs as designed, it offers much better control in comparison with thermal annealing or a previous approach based on the combination of two different types of GNR precursors.[73]

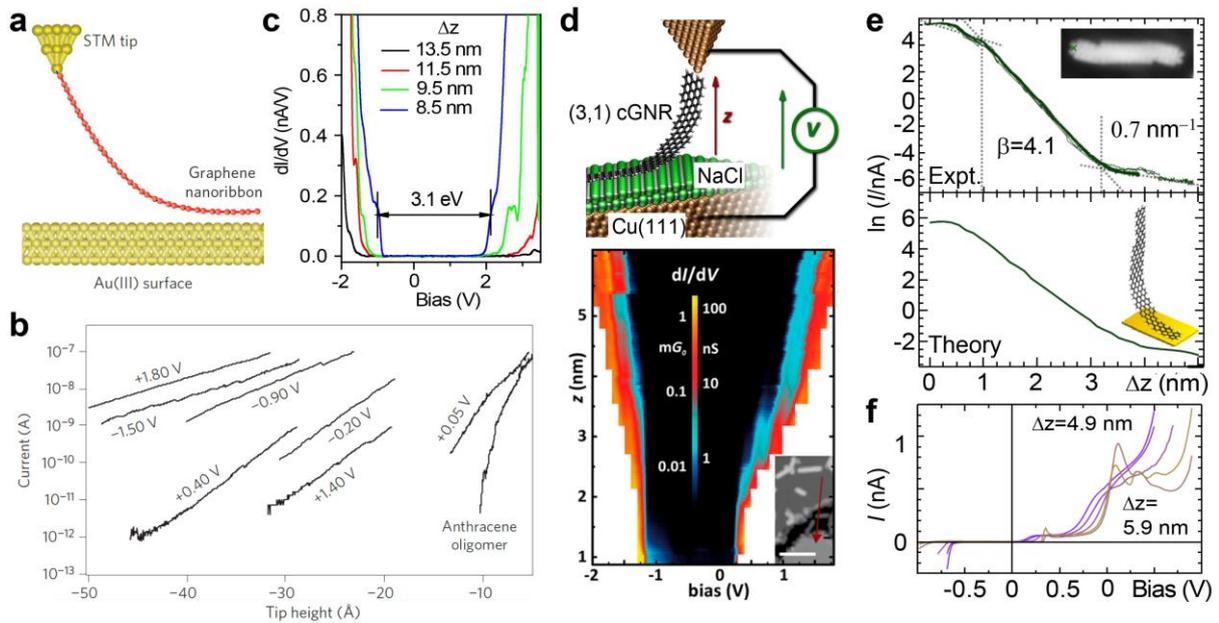

**Figure 15.** Transport through single ribbons measured by lifting with an STM tip. a) Configuration of the STM tip and the nanoribbon during a pulling sequence for a 7-aGNR. b) Current as a function of the tip height Δz for various biases (zero tip height refers to tip-surface contact). Panels (a,b) are reprinted with permission.[19b] Copyright 2017, Springer Nature. c) dI/dV spectra of the 7-aGNR at different pulling height, showing height-dependent bandgaps. d) Conductance mapping of (3,1)-chiral GNR on NaCl/Cu(111) (adapted with permission.[74] Copyright 2017, American Chemical Society). e) Experimental (top) and simulated (bottom) log current *vs*. Δz curves obtained while lifting the 5/7/5-GNR(2,4,2) HJ (STM image shown in inset). f) Experimental *I-V* curves recorded at different tip height Δz for the 5/7/5-GNR(2,4,2) HJ. Panels (e,f) are reprinted with permission.[75] Copyright 2017, Springer Nature, licensed under CC BY 4.0.

Besides fabricating GNRs and HJs, the STM also allows testing of the designed device functions, particularly their transport properties. Short channel field-effect transistor (FET) devices with a channel length of about 20 nm were demonstrated for the 7, 9 and 13-aGNRs.[76] However, it is extremely hard to place exactly one GNR in between lithographically fabricated contact electrodes. To access the transport properties of individual GNRs, an STM break junction method can be used, where single GNR or molecule bridges the tip and the substrate to allow two-point contact measurements, as demonstrated for polymer chains.[19a] Koch et al. applied the method to the 7-aGNRs and demonstrated that GNRs could be lifted up by using an STM tip attaching to the short zigzag end, as illustrated in **Figure 15**a.[19b] Bias-dependent conductance of single 7-aGNRs was measured in the lifting configuration (Figure 15b). The exponential dependence of current on the tip height indicated a tunneling conductance in the GNR. d$I$/d$V$ spectra showed a height-dependent band gap for the 7-aGNRs (Figure 15c). The measured band gap increases as the GNR is lifted higher, and approaches the GW calculated value (3.7 eV) for an isolated 7-aGNR.[77] A height-dependent band gap was also observed in a GNR with chiral edges (Figure 15d),[74] confirming the reduction of substrate screening as GNRs are lifted up from the surface.

The STM break junction method was also used to measure the transport properties of GNR HJs. Jacobse et al. reported the transport properties of GNR HJs made of 5 and 7-aGNRs (Figure 15e).[75] The 5-aGNRs have a smaller gap than the 7-aGNRs,[78] which reduces the Schottky barrier when the 5-aGNRs are used in between the STM tip and the 7-aGNRs.[66b] At the increased lifting height, the NDR was observed in the 5/7/5-aGNR(2,4,2) HJ, formed with 5, 7 and 5-aGNR segments with 2, 4 and 2 unit cells, respectively (Figure 15f).

## 6. Writing Atomic Scale Devices on Semiconductor Surfaces

The application of STM-based atomic scale manipulation on semiconducting surfaces was proposed in the early 1990s to realize atomic-scale functional systems.[5] Particular attention was given to the (001) face of group IV semiconductors, such as Si(001) and Ge(001).[79] As the Si and Ge are fully compatible with the complementary metal-oxide semiconductor (CMOS) technologies, the fabricated systems are readily adaptable for device applications.[79c]

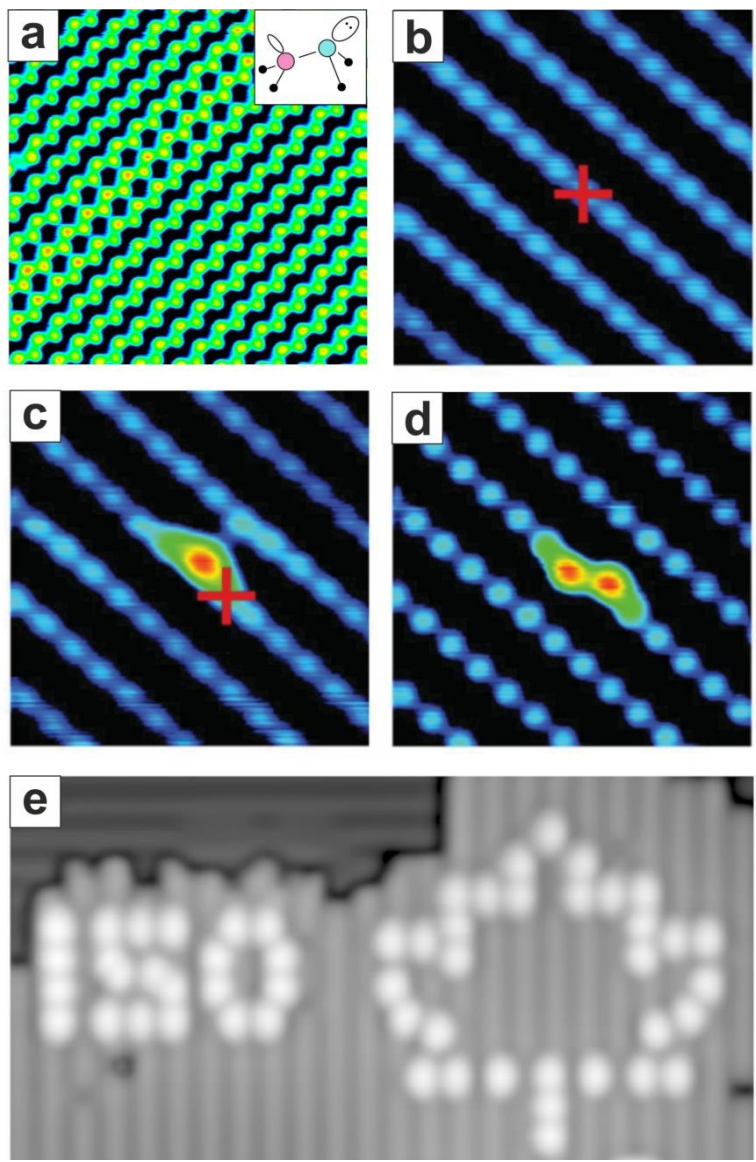

**Figure 16**. a) Occupied state STM image of a bare Ge(001) surface obtained at cryogenic temperatures. Germanium dimer rows forming p(2×2) and c(4×2) reconstruction areas are clearly seen. Inset shows a schematic presentation of a buckled dimer possessing dangling bonds (DBs). b-d) Series of STM images presenting sequential dimer-by-dimer formation of a DB structure: fully hydrogenated surface area (b), single DB dimer (c) and two DB dimers (d). Crosses mark the positions where current-voltage pulses are applied. Panels (b-d) are adapted with permission.[80] Copyright 2012, American Physical Society. e) Atomically-precise DB pattern for the 150th anniversary of Canadian Confederation fabricated by automatized sequential desorption of single H atoms (reprinted with permission.[81] Copyright 2018, Springer Nature, licensed under CC BY 4.0).

The surfaces of Si(001) and Ge(001) are a highly attractive platform for atomic-scale manipulation. First, its preparation method under UHV conditions is well developed to form reconstruction surfaces containing rows of dimerized atoms possessing unsaturated bonds (**Figure 16**a).[82] These surface dangling bonds (DBs) host well-defined electronic states, which are the key elements for any implemented functionality on top. Second, the surface atoms of Si(001) or Ge(001) possessing DBs are chemically active, so these surfaces are easily passivated by adsorbates that saturates DBs.[83] The most prominent examples are hydrogen passivated Si(001):H or Ge(001):H surfaces, where each DB is saturated by a single hydrogen atom (Figure 16b).[83b, 84] Third, hydrogen atoms on Si(001):H and Ge(001):H surfaces can be selectively removed by applying a current-voltage pulse from the STM tip (Figure 16c-d).[83b, 85] It makes these surfaces an ideal platform for STM-based lithography. Recent progress on experimental protocols and STM instrumentation are enabling the automatic fabrication of the atomic structures in truly atom-by-atom [81, 86] or dimer-by dimer [80, 86b, 87] fashion (Figure 16e). Below, we will describe several examples of functional atomic systems implemented on the Si(001 and Ge(001) surfaces by atomically precise manipulation with STM. In all cases, the structures are extremely fragile to external contaminations and thus testing their functionalities requires *in situ* characterization. The recent development of the multiprobe scanning tunneling microscopy/spectroscopy (MP-STM/STS) makes it possible to directly characterize the electronic transport of these systems.

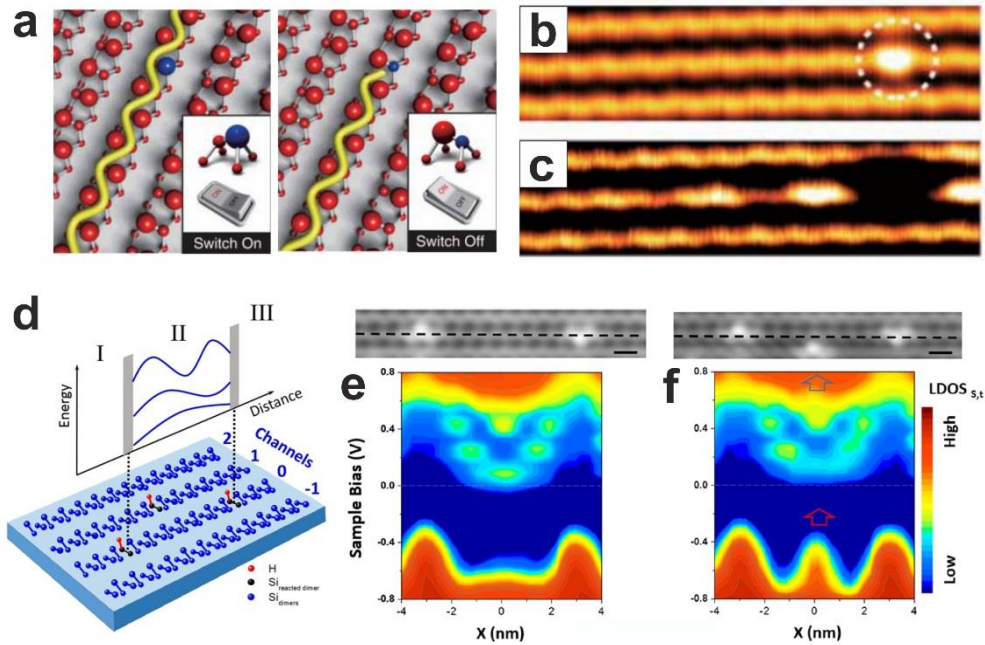

**Figure 17.** Atomic-scale functional systems on bare surfaces of Ge(001) and Si(001). a-c) Atomic switch composed of the Sn-Ge dimer on a Ge(001) surface. a) Schematics of propagation of electrons around the U (left) and L (right) positions of atomic-switch. The red and blue atoms represent Ge and Sn atoms, respectively. The yellow curve shows a conduction pathway for the electron in π* electronic states. b,c) The unoccupied state d$I$/d$V$ maps of U and L switch configuration, respectively. Panels (a-c) are adapted with permission.[88] Copyright 2007, AAAS. d-f) Coherent QID implemented on a bare Si(001) surface. d) Schematics of QID showing the Si dimer rows (blue) together with the quantum well (channel 0) formed between two hydrogen atoms (red) located on two dimers (black). e) d$I$/d$V$ map of a quantum well formed by two H atoms adsorbed on a Si dimer row. f) d$I$/d$V$ map of a quantum well gated by additional hydrogen atom located on the adjacent Si dimer row (corresponding STM images are presented on the top of d$I$/d$V$ maps). Panels (d-f) are adapted with permission.[89] Copyright 2015, American Chemical Society.

Manipulation over surface atoms on the bare surfaces of Si(001) and Ge(001) exhibited a proof-of-principle demonstration of the device functionality from DB-induced surface states.

The bare (001) surfaces of Si and Ge displays (2×1) surface reconstructions that are fully covered with DBs associated with each surface atom. DBs create electronic bands that are decoupled from the bulk states of the substrates. These DB-induced surface bands are highly anisotropic due to the interaction between the Si/Ge atoms within the surface dimer reconstruction rows. As a result, the DB-bands are strongly dispersive along the [110] direction of the rows and the system can be treated as quasi-1D.[39, 88-90] Tomatsu et al. showed that electronic transport of charge carriers within such a single dimer wire on Ge(001) could be controlled by an atomic switch operated by STM.[88] The small coverage of Sn atoms deposited at room temperature is incorporated into Ge(001) lattice by substitution of single Ge atoms in buckled dimer reconstruction wires (**Figure 17**a). The authors proposed to use the Sn-Ge dimer as a functional switch. Upon STM *I-V* pulses the buckling geometry of Sn-Ge dimer is reversibly changed between two tilted configurations. In this way, Sn atoms could be controllably positioned either in upper (U) or lower (L) configuration with respect to Ge atom from the dimer. Interestingly, the exact configurations of the switch had severe consequences to the electronic transport within the dimer wire. The authors reported that for U switch configuration the electronic transport is not affected, while the L configuration effectively backscatters coherent electronic carriers. The claim is proved indirectly by STM d*I*/d*V* maps, which showed characteristic modulations in LDOS related to inference between incoming and backscattered quasiparticle wavefunctions for L but not for U configuration (Figure 17b,c). Due to the considerable electronic decoupling from the bulk, coherence length of the charge carriers in quasi-1D wires reaches tens of nanometers at low temperature.[39, 91] These characteristics make the surfaces a good candidate for the implementation of prototypical device functionalities based on quantum coherence of the electronic transport. Naydenov et al. took this advantage and proposed the realization of quantum interference device (QID) on n-type doped bare Si(001) surface.[89] The exact atomic structure of the QID consists of two single hydrogen atoms adsorbed on the two Si dimers at the same reconstruction row (Figure 17d). Because the Si substrate is n-type, the resulting single DB sites formed next to the adsorbed hydrogen atoms are doubly occupied by electrons and thus host a net negative charge. The H atoms separated by a few nm creates two potential barriers, forming a quantum well (QW) for unoccupied electronic states of the dimer wire in the central scattering region (Figure 17e). Importantly, external regions are coupled to these QW states through the barriers. These couplings are responsible for QID operation, where energy-dependent electron transport through the QID is governed by the exact states generated within the central QID scattering region. The authors propose to operate the QID by engineering energy position of the QW states through electrostatic gating. For example, they placed or removed additional H atoms next to QID scattering region, then negative net charge related to the single DB site formed additional potential step within QID. In figure 17f, a DB site is located on the neighboring reconstruction row, and LDOS shows the shift in the energy of QW states induced by the potential step from the neighboring DB. The energy-dependent electronic transport *I-V* characteristics of QID are expected to shift accordingly, implementing intended device functionality.

Even higher level of control over the exact electronic structure of a functional system is possible with the hydrogenated Si(001):H and Ge(001):H surfaces. The monohydride (2×1) faces of Si(001):H or Ge(001):H surfaces are obtained by exposing the surfaces to atomic hydrogens.[83b, 84] The local defects in the form of single or dimer DBs provide contrasts in chemical reactivity and electronic states. The chemical reactivity of the DBs is much higher than the hydrogenated parts. Similarly, the adsorbates on the hydrogenated surfaces are electronically decoupled from the substrate while the DBs cause strong chemical bonds/interaction between the adsorbates and the substrates.[92] The decoupling by hydrogen layer was shown to preserve the electronic structure of prototypical molecules,[92a, 92b, 92d] which could be used for the implementation of single molecule-based electronic functional devices.[54d, 83d, 93] On the other hand, the formation of chemical bonds between DBs and adsorbates was applied for a variety of applications including epitaxial growth,[83a, 94] precise crystal doping [79c, 95] or controlled interaction with large organic molecules.[54d, 83c, 83d, 93, 96] The single or dimer DBs induces localized defect states whose coupling can be precisely controlled by their spatial distribution.[80, 86, 97] This fact in combination with selective removal of hydrogen atoms by STM-based lithography provides a unique platform for implantation of functional systems. In the seminal work by Shen et al., authors discussed two hydrogen desorption regimes depending on the bias voltage applied between the sample and the tip.[84b] For the sample bias exceeding 5 V, field-emitted electrons cause desorption of hydrogen atoms with an effective resolution of several nm depending on the tip morphology.[98] This is particularly useful for patterning the hydrogen layer in hundreds of nanometer scales.[83b, 85] More importantly, in the tunneling current regime with the bias about 2 ~ 4 V, inelastically scattered electrons emitted from STM-tip desorb hydrogen atoms through multiple vibrational excitation mechanism.[84b, 84c] Due to much lower yield and high-order dependence on tunneling current that is strongly localized under the exact STM apex, this process of desorption reaches the atomic precision. In either way, the patterned bare surface atoms exhibit chemical activity towards molecular adsorbates that has numerous applications.[79c, 94, 99] Additionally, room temperature STM lithography and corresponding thermal stability of atomically precise DB patterns is one of the key advantages of these systems.[79a, 97h, 99-100] However, nowadays STMs operating at cryogenic temperatures have the picometer stability for the tip positioning without feedback, which significantly enhances the reliability of lithographic processes by tracking tunneling current *vs.* time spectra during bias pulse and subsequent registration of single H desorption events.[16, 80-81, 86a, 87] Finally, STM-based manipulation can be used for the reverse process, where hydrogen atom can be deposited on a DB-site from the non-contact AFM[101] or STM tip.[81] The method provides an error-correction tool that is essential for scaling up the capabilities of atomically precise STM lithography.

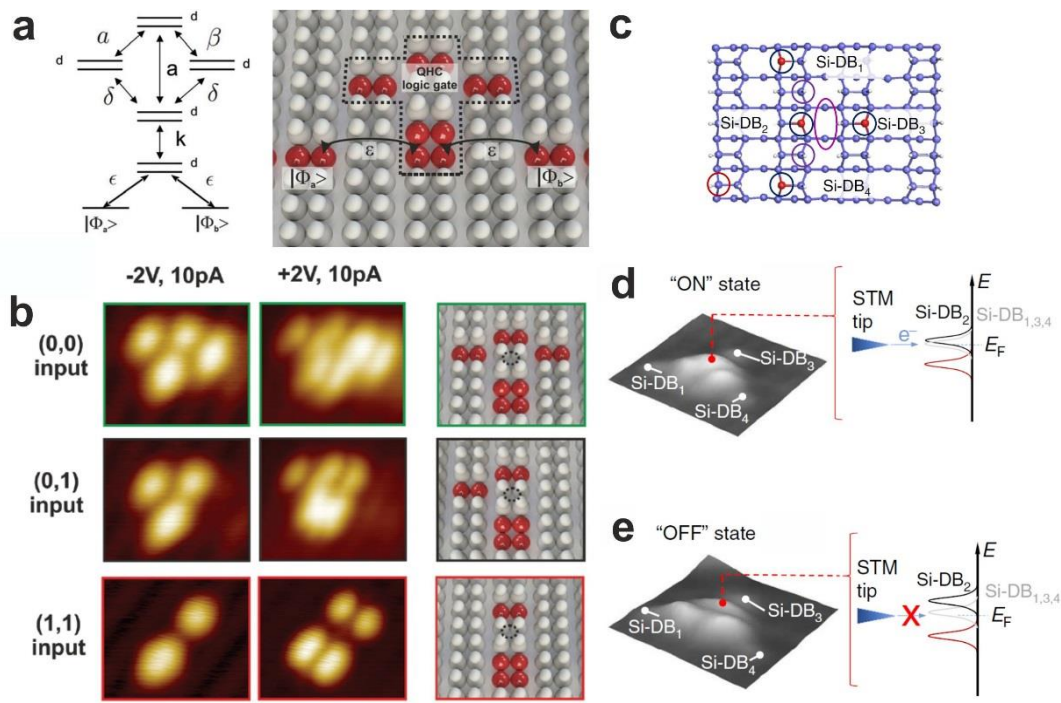

**Figure 18.** Functional systems on hydrogenated Si(001):H surface. a,b) Quantum Hamiltonian Boolean (QHC) logic gate. a) The schematic of the simplified quantum system including the NOR/OR calculating QHC circuit and the $|\phi_a\rangle$ and $|\phi_b\rangle$ output reading states. In the planar implantation of the NOR/OR QHC logic gate the red and grey balls represent bare and hydrogenated Si atoms, respectively. b) The realization of dangling bond NOR/OR logic gates on a Si(001):H surface. The logical inputs for the different structures of the gate are given on the left side of the STM images. Panels (a,b) are adapted with permission.[97c] Copyright 2015, Royal Society of Chemistry. (c-e) ON/OFF switching device. c) Model structure of the device composed of four Si dangling bonds. d,e) Three-dimensional STM topographic images of two switchable configurations of the device with schematics presenting the proposed implementation of its functionality. Panels (c-e) are adapted with permission.[97e] Copyright 2017, Springer Nature, licensed under CC BY 4.0.

To demonstrate the device functionality of the patterned DBs on hydrogenated surfaces, various types of atomic-scale logic gates were proposed and fabricated. Kolmer et al. used the relative electronic couplings between DB dimers on Si(001):H to implement prototypical

Boolean logic gate structure.[97c] The principle of device operation is based on Quantum Hamiltonian Computing (QHC) approach.[102] The designed DB array is described by $H_0$ Hamiltonian matrix consisting of coupling elements between ten DBs (**Figure 18**a). The classical inputs of the QHC logic gate should change the chosen $H_0$ matrix elements α and β that affects the eigenstate spectrum. The inputs are realized by removing (logic input 0) or adding (logic input 1) hydrogen atoms at two input Si dimers (Figure 18b). The readout of the Boolean truth table is then performed by tracking displacement of the well-defined $H_0$ eigenstates by two external readout degenerate states $|\phi_a\rangle$ and $|\phi_b\rangle$ which represent two external interconnects. If some of $H_0(\alpha,\beta)$ eigenstates resonates with $|\phi_a\rangle$ and $|\phi_b\rangle$ for chosen energy, the quantum system exhibits fast Heisenberg-Rabi oscillations between these readout states. These fast oscillations are then reflected in increased tunneling current registered between two interconnects. In the prototypical experiment with single-probe STM, two planar interconnects are substituted by a vertical configuration where $|\phi_a\rangle$ and $|\phi_b\rangle$ are represented by STM tip and Si substrate, respectively. In this case, QHC logic gate NOR and OR operations are tracked by d$I$/d$V$ spectra taken at different QHC gate structures.

Another approach was proposed by Yengui et al., who used STM to fabricate functional structure composed of four single DBs forming Y shaped geometry on Si(001):H surface (Figure 18c).[97e] The structure was switched reversibly between negative and neutral charge states by STM bias $I$-$V$ pulses. Due to the Jahn-Teller like distortion, these charge transitions are accompanied by reorganization of the Si lattice. As a result, switching by bias pulses changes the energy of the electronic states of four DB structure (Figure 18d,e). The authors used defined electronic states from the central part of the device to implement ON/OFF conduction switch. The intended device operation was confirmed by tracking d$I$/d$V$ resonances observed in Si(001):H band-gap energy region by STS.

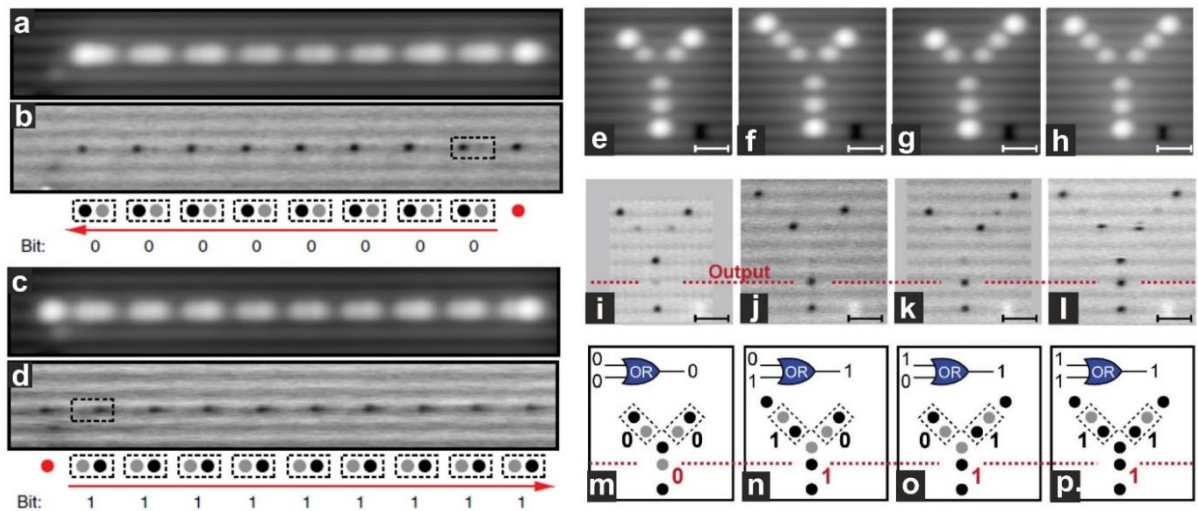

**Figure 19.** Implementation of binary logic elements in dangling bond (DB) structures on Si(001):H surface. a-d) Information transmission through a DB binary wire. Filled states STM image (a) and corresponding constant-height zero bias Δf image (b) of an eight-pair wire with a non-paired gating single DB (red circle) on the right. STM image (c) and constant-height zero bias Δf image (d) of a nine-pair wire after adding a perturbing DB (red circle) on the left. Guides are placed below b and d to show the location and bit state of the pairs. e-p) OR gate constructed of dangling bonds. e-h) Filled state STM images of the OR gate in various actuation states. i-l) Corresponding constant-height zero bias Δf images of the gate, showing electron locations as the dark depressions, with the output marked in red. m-p) Models of the gate presenting the binary status of each unit. Scale bars are 2nm. Adapted with permission.[97g] Copyright 2018, Springer Nature.

Wolkow group developed an alternative approach for realizing logic gates based on the quantum cellular automata (QCA) concept (**Figure 19**).[79a, 97h] In this case, localized states of isolated DBs act as quantum dots. In properly designed DB structures, these quantum dots encode single information bits which are coupled together so that transmission of binary information through the system is possible in an up-scaling manner. In their recent work, Huff et al. proposed and experimentally demonstrated the full strategy for implementing QCA in silicon DBs on Si(001):H surface.[97g] First, they encoded information bit in a pair of single DBs located on neighboring dimers. These DB pairs at n-type doped Si substrate share a single negative charge, whose distribution could be manipulated between two DB sites by external electrostatic gating. The gating could be realized by biased STM or NC-AFM tips,[97b, 97d, 97f, 103] [97g] or by nanostructures fabricated on the surface.[104] For example, negatively charged single DB in the vicinity of a DB pair breaks the symmetry in charge distribution within the DB pair. The authors proposed to encode a single bit of information in such asymmetric charge distribution within one DB pair. Moreover, the asymmetric charge distribution in one pair could be used to gate neighboring pairs and thus providing intended coupling between quantum dots crucial for QCA operation. The transmission of binary information within a wire consisting of 9 DB pairs is shown in Figure 19a-d. The structure is fabricated by STM-lithographic protocols,[81] whereas the charge states of DBs are determined by zero bias constant height NC-AFM images. The latter approach with NC-AFM is crucial, because biased STM tip is known to dynamically affect the charge states of DBs.[97b, 97d, 97f, 103] In a similar manner, Huff et al. also demonstrated a basic three terminal DB structure processing binary information encoded in quantum dots as an OR logic gate (Figure 19e-p). So far, all discussed functional atomic-scale systems were fabricated and *in situ* operated by single probe SPM. Thus, the functional properties were determined indirectly from experiments combined with theoretical modeling studies. As a result, all the anticipated control over electronic transport properties and thus related information transmission and processing within atomic-scale systems have not been tested directly in a more practical planar configuration. To probe information transmission through an atomic-scale functional system, it must be contacted with atomic-scale precision by at least two interconnects. Such precision is beyond the capabilities of conventional top-down lithographic processes. Even the STM-based lithographic strategies leading to formation of highly P- or B-doped metallic wires and conductive contacts on Si(001) and Ge(001) surfaces cannot be straightforwardly combined with maintaining chemically active DB structures on the surface, which would be in this case easily saturated, because the processes require substrate encapsulation and further thermal activation of dopants [79c, 95a, 95b].

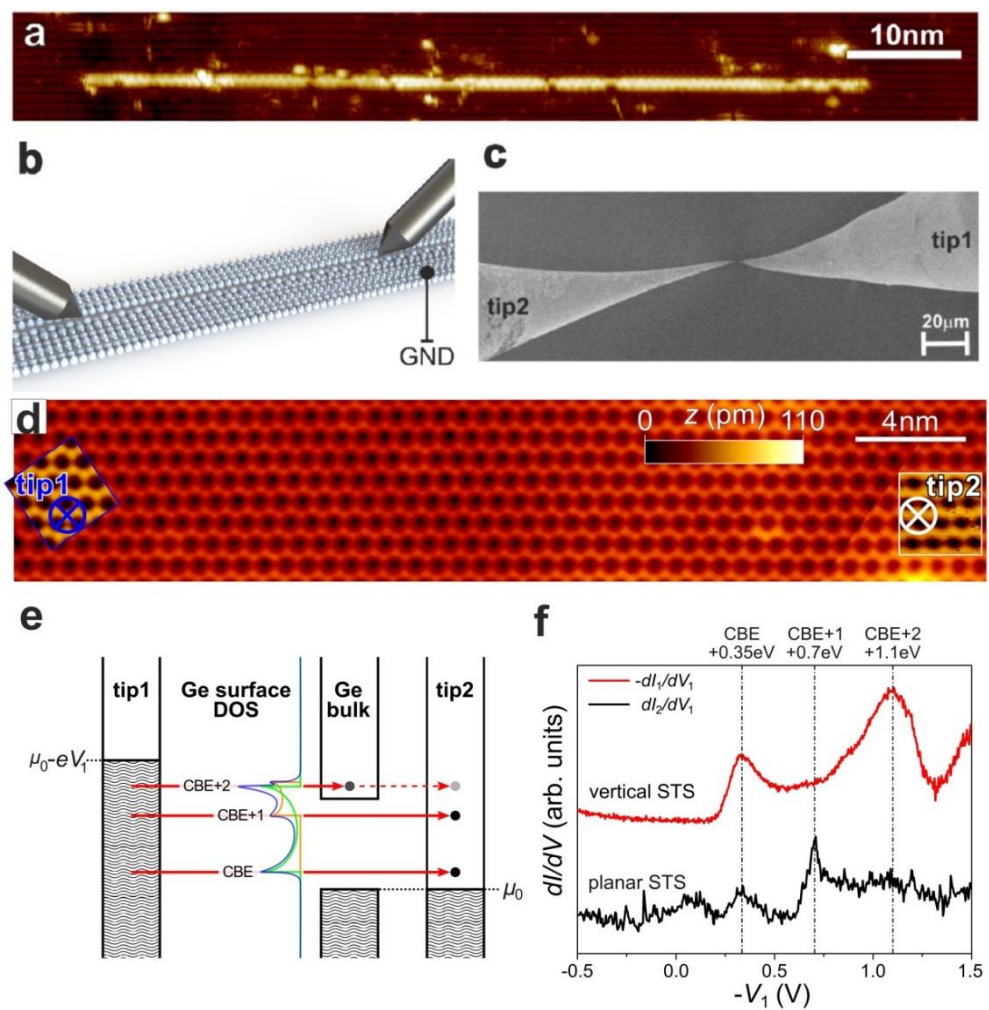

**Figure 20.** Two-probe STM/STS experiments. a) STM image of a DB dimer wire on the Ge(001):H surface. About 70 nm long wire consists of 156 bare Ge dimers (DB dimers) and has 14 atomic scale defects. The structure is constructed by several line desorption procedures (feedback loop closed) followed by a few single hydrogen atoms desorption procedures (feedback loop open). b) Schematic view of the two-probe experiment geometry. Both STM probes approach the same atomic-scale wire of bare Ge dimers along Ge(001):H reconstruction rows. c) SEM image of two STM probes approached to Ge(001):H surface. Panels (a-c) are adapted with permission.[16] Copyright 2017, IOP Publishing. d) STM images

of the c(4×2) reconstructed Ge(001) surface obtained prior to the 2P-STS experiment. Insets show two atomically resolved STM images obtained simultaneously by both probes. STM probe positions for 2P-STS are marked by blue and white circles. e) Two-probe measurement scheme, which probes the energy positions of ballistic transport channels mediated by the surface states. The Ge dimer wire density of states shows the three resonances, associated with the edges of quasi-1D bands. Note that limited band-gap of the bulk Ge could affect transconductance signal measured for CBE+2 resonance. f) Vertical (standard) $dI_1/dV_1$ and planar transconductance $dI_2/dV_1$ 2P-STS signals as a function of tip 1 voltage. The resonances observed in the $dI/dV$ characteristics at energies 0.35 eV, 0.7 eV and 1.1 eV are ascribed to the CBE, CBE+1 and CBE+2 resonances shown schematically in (e). Panels (d-f) are adapted with permission.[39] Copyright 2019, Springer Nature.

An attractive, yet challenging alternative to overcome these obstacles is to use multiprobe STM for *in-situ* characterization of electronic transport or information transmission in fabricated functional devices.[30d] As reported by Kolmer et al., current *state-of-the-art* cryogenic multiprobe STM can reach the expected atomic-scale precision in contacting nanoscale systems.[16] The authors presented the full methodology behind such experiments on the example of about 70nm long DB dimer wire fabricated on a hydrogenated Ge(001):H surface by STM-lithography (**Figure 20**a). The wire was approached by two STM probes, which independently scan the surface in tunneling conditions (Figure 20b). These two macroscopic probes were initially approached to about 1μm separation with the use of scanning electron microscope (SEM) navigation (Figure 20c). After this initial stage, the whole experiment fully relied on STM imaging, which enables atomic-scale lateral positioning of the probes. Moreover, about 2 pm stability in tip-to-surface distance allows precise determination of contact resistance ranging from purely tunneling to single atomic contact regimes. The stability provides reliable conditions for atom manipulation. This is admittedly demonstrated by the DB wire formation during the two-probe experiment. Importantly, *I-V* pulses used during STM-lithography did not affect the shapes of both STM apexes. This was proved by the simultaneous approach to the patterned wire with the probe-to-probe separation distance reaching about 30 nm. Such configuration allows two-probe *I-V* characterization of the system. The *I-V* characteristics showed pronounced resonances for electron energies reaching unoccupied DB-related bands [80].

To go beyond this achievement and reveal the exact relationship between the electronic band structure of a system and the multiprobe transport results, atomically precise model systems at tens of nanometer scales are desirable. In their recent work, Kolmer et al. used quasi-1D electronic structure of unoccupied states of the bare Ge(001) surface to establish a protocol for two-probe STS (2P-STS).[39] Applying the methodology presented in the previous work, the STM probes were simultaneously position at the very same Ge dimer wire about 30 nm apart as presented in Figure 20d. To extract the detailed information about energy-dependent electron transport properties of this system, the authors proposed the following 2P-STS experimental design. Both probes are kept in constant height above the grounded sample, which defines the respective energy of characterized surface electronic states (see scheme in Figure 20e). Bias $V_1$ is applied to *tip 1*, which injects hot charge carriers into the electronic states of the system. This source probe is kept in tunneling condition with large resistance. Second probe that is virtually grounded acts as a collector. If the mean free path of the hot charge carriers is longer than the probe-to-probe separation, some of the injected carriers propagate coherently through the Ge dimer wire, and then they are registered by *tip 2* as $I_2$ signal. Thus, the 2P-STS transconductance signal, defined as d$I_2$/d$V_1$, carries the energy-dependent information about transport properties of the atomic-scale system. This was proved by showing the relationship between transconductance resonances and the electronic band structure of the bare Ge dimer wire in function of injected carriers' energy (Figure 20f). The results comprise fundamental steps towards the complete realization of planar, atomic-scale electronic prototypical functional devices.

## 7. Perspective and Conclusion

Further development of STM-based atomic scale manipulation calls for solutions to two major obstacles. First, it is necessary to develop autonomous fabrication methods capable of creating large and complex structures with atomic precision. Second, it is necessary to protect the fabricated systems and make them compatible with ambient environment while retaining the designed functionalities.

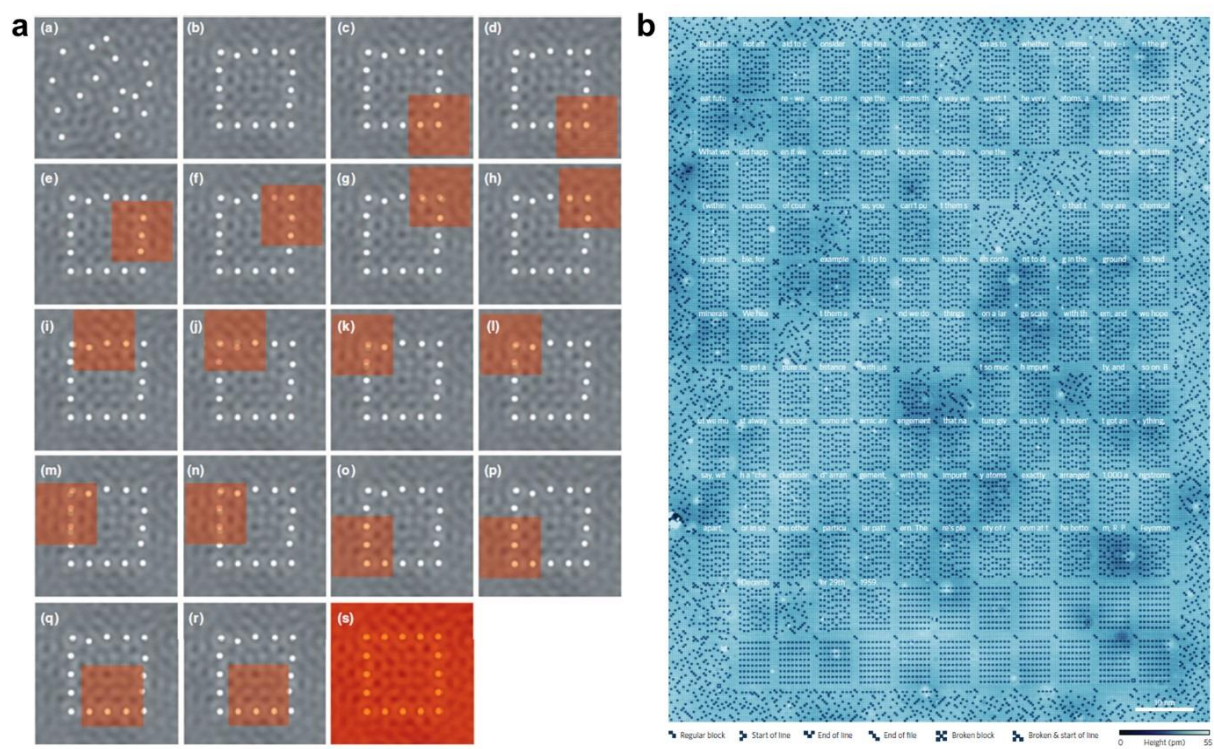

**Figure 21.** Autonomous assembly of atomic structures with STM. a) Sequence of topographic images showing autonomous assembly of Co atoms on Cu(111) from randomly placed positions to perfect square (reprinted with permission.[105] Copyright 2014, AIP Publishing). b) 1,016-byte atomic memory realized on Cl/Cu(100) with autonomous assembly (reprinted with permission.[18a] Copyright 2016, Springer Nature).

Now there are various attempts to automate the whole process of atomic-scale manipulation including detection of atoms, determination of manipulation path, assembly and confirmation of final structure.[105] **Figure 21**a shows the autonomous assembly of a simple square with Co atoms on Cu(111) by controlling STM with an automatic protocol. First, the software calculated the trajectory for assembling randomly distributed Co atoms to the defined square, and lateral manipulation was performed along the trajectories which brought the atoms roughly near the defined square. Then, to correct the errors, the automatic program sectioned the structure to the smaller parts (red squares in Figure 21a) and went over each part sequentially to calculate the new trajectories and do the lateral manipulation so that all the atoms are in the exact positions defined by the user. This approach was extended to other substrates and larger-scale structures. Kalff et al. realized a kilobyte atomic memory by manipulation of Cl vacancy defects on Cl deposited Cu(100) (Figure 21b).[18a] Achal et al. realized 192-bit atomic memory by patterning DBs with STM lithography on the Si(001):H surface.[81] In both cases, the manipulation process was automatized with the software detecting atoms, calculating trajectories, and performing error corrections. These demonstrations showed the proof-of-principles of autonomous fabrication of atomic-scale structures in commercially available STM instruments. Further improvements in image recognition can leverage the development of machine learning algorithms.[106]

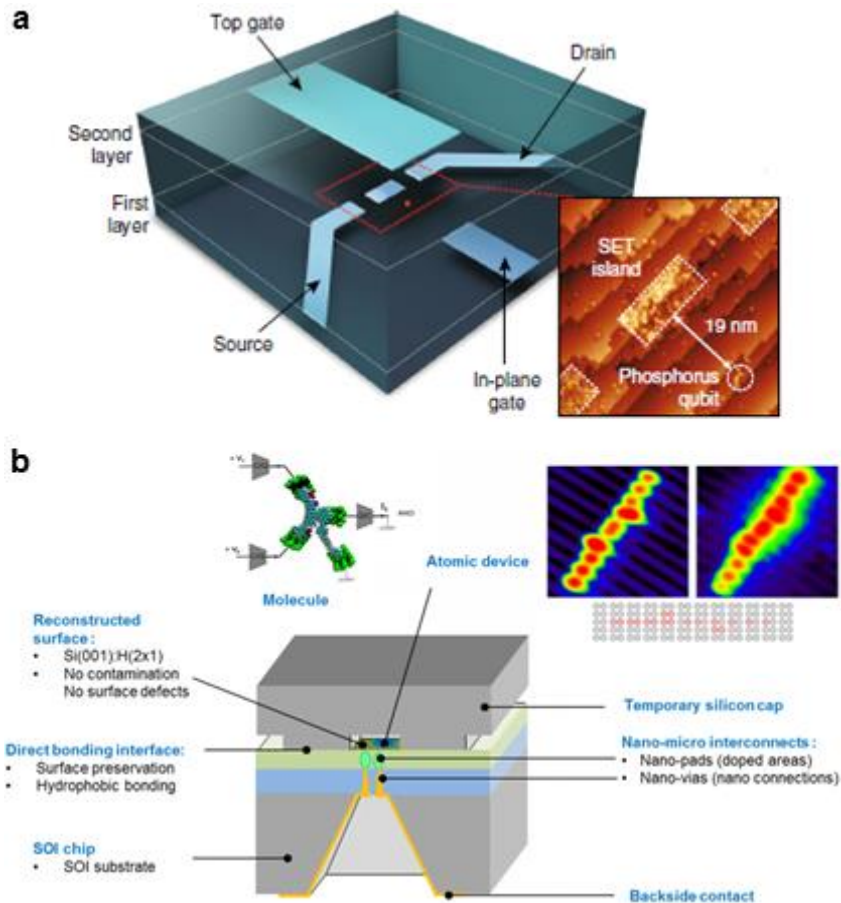

**Figure 22.** Capping of atomic structures for *ex situ* device application. a) Atomic device made with STM lithography. Phosphorus donors were implanted in atomically precise positions, and then capped with epitaxial Si layers together with the metal contacts for source, drain and gate electrodes (reprinted with permission.[107] Copyright 2019, Springer Nature). b) Schematic of a microelectronic chip allowing connection of an atomic-scale device realized on the functionalized Si(001):H surface. A temporary Si cap containing a cavity provides the hermetic seal and ensures the nano-object protection. The connection from the object at the nanoscale is ensured by nano–micro interconnects toward the backside of the chip. Insets are STM images of a DB structure fabricated on the samples sealed with temporary Si cap and then de-bonded in the UHV chamber (adapted with permission from chapter 2 in ref. [79b]. Copyright 2017, Springer Nature. See also ref. [79c]).

Atomically precise structures are typically fabricated at low-temperature and under UHV conditions for clean surfaces and thermal stability. To take the fabricated structures into the ambient environment, additional protection steps are required. One example is capping the structures with appropriate protecting overlayers. In this respect, applications of STM-formed lithographic patterns on hydrogenated Si(100) or Ge(001) surfaces have made great advancements. By dosing phosphine gas, phosphorus atoms can be absorbed in atomically defined positions. Then, the structure is capped by a few nanometer thick Si overlayers on top and dopants are activated by thermal annealing (**Figure 22**a).[11, 18b, 24, 95a, 107] Further lithographic processes can be applied to fabricate interconnects. Following this procedure, functional devices with atomically precisely positioned dopants have been demonstrated, including single atom transistors, a-few-atom-wide nanowire, and spin qubits.

Figure 22b shows another method for protecting the atomic scale functional systems. Here, the structures were patterned on the hydrogenated Si(001):H surface and protected by directly bonding with another hydrogenated Si(001):H (temporary Si cap in Figure 22b).[86a] In such a nanopackage, the fabricated atomic structures are sealed in small cavities between two Si wafers. This method can also be used to protect non-patterned Si(001):H surfaces so that the wafers can be stored in ambient condition and be de-bonded under UHV conditions again for later applications. The de-bonded Si(001):H surface showed very low defect density without the need of additional UHV cleaning protocols.[86a] By combining with the strategies of contacting atomic-scale systems on Si(001) or Ge(001) surfaces,[79c, 95a, 95b, 104, 108] the nanopackaging concept brings the functional quantum devices made by atomic scale manipulations close to the practical applications.

In conclusion, the continuous progress in atomic scale manipulations with STM has demonstrated a wide range of applicability, from metal to semiconductor surfaces, from single to conjugated molecules, and from surfaces to subsurfaces. The artificial atomic structures have revealed exotic physical properties that otherwise only resided in theoretical predictions.

Precise control over atomic structures and their environment provides unprecedentedly defined conditions for accurate experimental validation of various condensed-matter theories, deepening our understanding of quantum physics. In parallel, the development of STM instrumentation and control systems in both hardware and software is constantly improving the scalability and reliability of this approach. Further progresses will bring these artificially created atomically precise structures into practical environment and make the approach more accessible for the realization of novel quantum architectures.

## Acknowledgments

This work was conducted at the Center for Nanophase Materials Sciences, which is sponsored at Oak Ridge National Laboratory by the Office of Basic Energy Sciences, U.S. Department of Energy.

## References

[1]     G. Binning, H. Rohrer, C. Gerber, E. Weibel, *Phys. Rev. Lett.* **1982**, 49, 57.
[2]     D. M. Eigler, E. K. Schweizer, *Nature* **1990**, 344, 524.
[3]     C. Tourney, *Engineering and Science* **2005**, 68, 16.
[4]     M. F. Crommie, C. P. Lutz, D. M. Eigler, *Science* **1993**, 262, 218.
[5]     J. W. Lyding, T. C. Shen, J. S. Hubacek, J. R. Tucker, G. C. Abeln, *Appl. Phys. Lett.* **1994**, 64, 2010.
[6]     L. Bartels, G. Meyer, K. H. Rieder, M. Wolf, G. Ertl, *Phys. Rev. Lett.* **1998**, 80, 2004.
[7]     S. W. Hla, L. Bartels, G. Meyer, K. H. Rieder, *Phys. Rev. Lett.* **2000**, 85, 2777.
[8]     N. Nilius, T. H. Wallis, W. Ho, *Science* **2002**, 297, 1853.
[9]     A. J. Heinrich, C. P. Lutz, J. A. Gupta, D. M. Eigler, *Science* **2002**, 298, 1381.
[10]    C. F. Hirjibehedin, C. P. Lutz, A. J. Heinrich, *Science* **2006**, 312, 1021.
[11]    M. Fuechsle, S. Mahapatra, F. A. Zwanenburg, M. Friesen, M. A. Eriksson, M. Y. Simmons, *Nat. Nanotechnol.* **2010**, 5, 502.
[12]    A. A. Khajetoorians, J. Wiebe, B. Chilian, R. Wiesendanger, *Science* **2011**, 332, 1062.
[13]    M. Peplow, *Nature* **2015**, 525, 18.
[14]    IBM,    A    Boy    And    His    Atom:    The    World's    Smallest    Movie, https://www.youtube.com/watch?v=oSCX78-8-q0, accessed: May, 2019.
[15]    D. Castelvecchi, *Nature* **2017**, 544, 278.
[16]    M. Kolmer, P. Olszowski, R. Zuzak, S. Godlewski, C. Joachim, M. Szymonski, *J. Phys. Condens. Mat.* **2017**, 29, 444004.
[17]    L. Bartels, G. Meyer, K. H. Rieder, *Appl. Phys. Lett.* **1997**, 71, 213.
[18]    a) F. E. Kalff, M. P. Rebergen, E. Fahrenfort, J. Girovsky, R. Toskovic, J. L. Lado, J. Fernández-Rossier, A. F. Otte, *Nat. Nanotechnol.* **2016**, 11, 926; b) B. Weber, S.

Mahapatra, H. Ryu, S. Lee, A. Fuhrer, T. C. G. Reusch, D. L. Thompson, W. C. T. Lee, G. Klimeck, L. C. L. Hollenberg, M. Y. Simmons, *Science* **2012**, 335, 64.

[19]   a) L. Lafferentz, F. Ample, H. Yu, S. Hecht, C. Joachim, L. Grill, *Science* **2009**, 323, 1193; b) M. Koch, F. Ample, C. Joachim, L. Grill, *Nat. Nanotechnol.* **2012**, 7, 713.

[20]   T. Kudernac, N. Ruangsupapichat, M. Parschau, B. MacIá, N. Katsonis, S. R. Harutyunyan, K. H. Ernst, B. L. Feringa, *Nature* **2011**, 479, 208.

[21]   K. K. Gomes, W. Mar, W. Ko, F. Guinea, H. C. Manoharan, *Nature* **2012**, 483, 306.

[22]   a) A. J. Heinrich, J. A. Gupta, C. P. Lutz, D. M. Eigler, *Science* **2004**, 306, 466; b) S. Loth, K. Von Bergmann, M. Ternes, A. F. Otte, C. P. Lutz, A. J. Heinrich, *Nat. Phys.* **2010**, 6, 340; c) D. Serrate, P. Ferriani, Y. Yoshida, S. W. Hla, M. Menzel, K. Von Bergmann, S. Heinze, A. Kubetzka, R. Wiesendanger, *Nat. Nanotechnol.* **2010**, 5, 350.

[23]   S. Loth, S. Baumann, C. P. Lutz, D. M. Eigler, A. J. Heinrich, *Science* **2012**, 335, 196.

[24]   M. Fuechsle, J. A. Miwa, S. Mahapatra, H. Ryu, S. Lee, O. Warschkow, L. C. L. Hollenberg, G. Klimeck, M. Y. Simmons, *Nat. Nanotechnol.* **2012**, 7, 242.

[25]   G. D. Nguyen, L. Liang, Q. Zou, M. Fu, A. D. Oyedele, B. G. Sumpter, Z. Liu, Z. Gai, K. Xiao, A.-P. Li, *Phys. Rev. Lett.* **2018**, 121, 086101.

[26]   C. Ma, Z. Xiao, H. Zhang, L. Liang, J. Huang, W. Lu, B. G. Sumpter, K. Hong, J. Bernholc, A.-P. Li, *Nat. Commun.* **2017**, 8, 14815.

[27]   a) L. C. Collins, T. G. Witte, R. Silverman, D. B. Green, K. K. Gomes, *Nat. Commun.* **2017**, 8, 15961; b) M. R. Slot, T. S. Gardenier, P. H. Jacobse, G. C. P. Van Miert, S. N. Kempkes, S. J. M. Zevenhuizen, C. M. Smith, D. Vanmaekelbergh, I. Swart, *Nat. Phys.* **2017**, 13, 672; c) S. N. Kempkes, M. R. Slot, S. E. Freeney, S. J. M. Zevenhuizen, D. Vanmaekelbergh, I. Swart, C. M. Smith, *Nat. Phys.* **2019**, 15, 127.

[28]   a) F. Ghahari, D. Walkup, C. Gutiérrez, J. F. Rodriguez-Nieva, Y. Zhao, J. Wyrick, F. D. Natterer, W. G. Cullen, K. Watanabe, T. Taniguchi, L. S. Levitov, N. B. Zhitenev, J. A. Stroscio, *Science* **2017**, 356, 845; b) J. Velasco, L. Ju, D. Wong, S. Kahn, J. Lee, H.-Z. Tsai, C. Germany, S. Wickenburg, J. Lu, T. Taniguchi, K. Watanabe, A. Zettl, F. Wang, M. F. Crommie, *Nano Lett.* **2016**, 16, 1620; c) Y. Wang, V. W. Brar, A. V. Shytov, Q. Wu, W. Regan, H.-Z. Tsai, A. Zettl, L. S. Levitov, M. F. Crommie, *Nat. Phys.* **2012**, 8, 653; d) Y. Wang, D. Wong, A. V. Shytov, V. W. Brar, S. Choi, Q. Wu, H.-Z. Tsai, W. Regan, A. Zettl, R. K. Kawakami, S. G. Louie, L. S. Levitov, M. F. Crommie, *Science* **2013**, 340, 734.

[29]   a) C. Ma, Z. Xiao, J. Huang, L. Liang, W. Lu, K. Hong, B. G. Sumpter, J. Bernholc, A.-P. Li, *Phys. Rev. Mater.* **2019**, 3, 016001; b) G. D. Nguyen, H.-Z. Tsai, A. A. Omrani, T. Marangoni, M. Wu, D. J. Rizzo, G. F. Rodgers, R. R. Cloke, R. A. Durr, Y. Sakai, F. Liou, A. S. Aikawa, J. R. Chelikowsky, S. G. Louie, F. R. Fischer, M. F. Crommie, *Nat. Nanotechnol.* **2017**, 12, 1077.

[30]   a) C. Durand, X. G. Zhang, S. M. Hus, C. Ma, M. A. McGuire, Y. Xu, H. Cao, I. Miotkowski, Y. P. Chen, A. P. Li, *Nano Lett.* **2016**, 16, 2213; b) S. M. Hus, X. G. Zhang, G. D. Nguyen, W. Ko, A. P. Baddorf, Y. P. Chen, A.-P. Li, *Phys. Rev. Lett.* **2017**, 119, 137202; c) W. Ko, G. D. Nguyen, H. Kim, J. S. Kim, X. G. Zhang, A.-P. Li, *Phys. Rev. Lett.* **2018**, 121, 176801; d) A. P. Li, K. W. Clark, X. G. Zhang, A. P. Baddorf, *Adv. Funct. Mater.* **2013**, 23, 2509.

[31]   a) S. W. Hla, *J. Vac. Sci. Technol. B* **2005**, 23, 1351; b) G. Meyer, L. Bartels, K. H. Rieder, *Comp. Mater. Sci.* **2001**, 20, 443; c) J. K. Gimzewski, C. Joachim, *Science* **1999**, 283, 1683; d) J. A. Stroscio, R. J. Celotta, *Science* **2004**, 306, 242; e) M. Ternes, C. P. Lutz, C. F. Hirjibehedin, F. J. Giessibl, A. J. Heinrich, *Science* **2008**, 319, 1066; f) T. C. Shen, C. Wang, G. C. Abeln, J. R. Tucker, J. W. Lyding, P. Avouris, R. E. Walkup, *Science* **1995**, 268, 1590.

[32]   B. C. Stipe, M. A. Rezaei, W. Ho, S. Gao, M. Persson, B. I. Lundqvist, *Phys. Rev. Lett.* **1997**, 78, 4410.

[33]   D. Wong, J. Velasco, Jr., L. Ju, J. Lee, S. Kahn, H. Z. Tsai, C. Germany, T. Taniguchi, K. Watanabe, A. Zettl, F. Wang, M. F. Crommie, *Nat. Nanotechnol.* **2015**, 10, 949.

[34]   J. Lee, D. Wong, J. Velasco Jr, J. F. Rodriguez-Nieva, S. Kahn, H.-Z. Tsai, T. Taniguchi, K. Watanabe, A. Zettl, F. Wang, L. S. Levitov, M. F. Crommie, *Nat. Phys.* **2016**, 12, 1032.

[35]   A. D. Oyedele, S. Yang, L. Liang, A. A. Puretzky, K. Wang, J. Zhang, P. Yu, P. R. Pudasaini, A. W. Ghosh, Z. Liu, C. M. Rouleau, B. G. Sumpter, M. F. Chisholm, W. Zhou, P. D. Rack, D. B. Geohegan, K. Xiao, *J. Am. Chem. Soc.* **2017**, 139, 14090.

[36]   G. A. Fiete, E. J. Heller, *Rev. Mod. Phys.* **2003**, 75, 933.

[37]   T.-H. Kim, Z. Wang, J. F. Wendelken, H. H. Weitering, W. Li, A.-P. Li, *Rev. Sci. Instrum.* **2007**, 78, 123701.

[38]   a) S. H. Ji, J. B. Hannon, R. M. Tromp, V. Perebeinos, J. Tersoff, F. M. Ross, *Nat. Mater.* **2012**, 11, 114; b) K. W. Clark, X. G. Zhang, I. V. Vlassiouk, G. He, R. M. Feenstra, A. P. Li, *ACS Nano* **2013**, 7, 7956.

[39]   M. Kolmer, P. Brandimarte, J. Lis, R. Zuzak, S. Godlewski, H. Kawai, A. Garcia-Lekue, N. Lorente, T. Frederiksen, C. Joachim, D. Sanchez-Portal, M. Szymonski, *Nat. Commun.* **2019**, 10, 1573.

[40]   a) W. Shockley, *Phys. Rev.* **1939**, 56, 317; b) S. D. Kevan, R. H. Gaylord, *Phys. Rev. B* **1987**, 36, 5809; c) P. M. Echenique, R. Berndt, E. V. Chulkov, T. Fauster, A. Goldmann, U. Höfer, *Surf. Sci. Rep.* **2004**, 52, 219.

[41]   a) K. F. Braun, K. H. Rieder, *Phys. Rev. Lett.* **2002**, 88, 968011; b) L. Bürgi, H. Brune, O. Jeandupeux, K. Kern, *J. Electron Spectrosc.* **2000**, 109, 33.

[42]   a) Y. Hasegawa, P. Avouris, *Phys. Rev. Lett.* **1993**, 71, 1071; b) M. F. Crommie, C. P. Lutz, D. M. Eigler, *Nature* **1993**, 363, 524.

[43]   H. C. Manoharan, C. P. Lutz, D. M. Eigler, *Nature* **2000**, 403, 512.

[44]   C. R. Moon, L. S. Mattos, B. K. Foster, G. Zeltzer, W. Ko, H. C. Manoharan, *Science* **2008**, 319, 782.

[45]   C. R. Moon, L. S. Mattos, B. K. Foster, G. Zeltzer, H. C. Manoharan, *Nat. Nanotechnol.* **2009**, 4, 167.

[46]   C. R. Moon, C. P. Lutz, H. C. Manoharan, *Nat. Phys.* **2008**, 4, 454.

[47]   F. Guinea, M. I. Katsnelson, A. K. Geim, *Nat. Phys.* **2010**, 6, 30.

[48]   M. Polini, F. Guinea, M. Lewenstein, H. C. Manoharan, V. Pellegrini, *Nat. Nanotechnol.* **2013**, 8, 625.

[49]   S. M. Hus, A. P. Li, *Prog. Surf. Sci.* **2017**, 92, 176.

[50]   D. Wong, *PhD Thesis*, University of California, Berkeley, **2017**.

[51]   J. Mao, Y. Jiang, D. Moldovan, G. Li, K. Watanabe, T. Taniguchi, M. R. Masir, F. M. Peeters, E. Y. Andrei, *Nat. Phys.* **2016**, 12, 545.

[52]   J. Velasco, J. Lee, D. Wong, S. Kahn, H.-Z. Tsai, J. Costello, T. Umeda, T. Taniguchi, K. Watanabe, A. Zettl, F. Wang, M. F. Crommie, *Nano Lett.* **2018**, 18, 5104.

[53]   a) K. Stokbro, C. Thirstrup, M. Sakurai, U. Quaade, B. Y.-K. Hu, F. Perez-Murano, F. Grey, *Phys. Rev. Lett.* **1998**, 80, 2618; b) J. Wyrick, X. Wang, P. Namboodiri, S. W. Schmucker, R. V. Kashid, R. M. Silver, *Nano Lett.* **2018**, 18, 7502; c) K. R. Rusimova, R. M. Purkiss, R. Howes, F. Lee, S. Crampin, P. A. Sloan, *Science* **2018**, 361, 1012.

[54]   a) S. Pan, Q. Fu, T. Huang, A. Zhao, B. Wang, Y. Luo, J. Yang, J. Hou, *P. Natl. Acad. Sci.* **2009**, 106, 15259; b) P. Liljeroth, J. Repp, G. Meyer, *Science* **2007**, 317, 1203; c) Q. Li, C. Han, M. Fuentes-Cabrera, H. Terrones, B. G. Sumpter, W. Lu, J. Bernholc, J. Yi, Z. Gai, A. P. Baddorf, P. Maksymovych, M. Pan, *ACS Nano* **2012**, 6, 9267; d) S. Godlewski, H. Kawai, M. Kolmer, R. Zuzak, A. M. Echavarren, C. Joachim, M. Szymonski, M. Saeys, *ACS Nano* **2016**, 10, 8499.

[55]   a) P. Maksymovych, D. B. Dougherty, X. Y. Zhu, J. T. Yates, *Phys. Rev. Lett.* **2007**, 99, 016101; b) N. A. Vinogradov, A. A. Zakharov, V. Kocevski, J. Rusz, K. A. Simonov,

O. Eriksson, A. Mikkelsen, E. Lundgren, A. S. Vinogradov, N. Mårtensson, A. B. Preobrajenski, *Phys. Rev. Lett.* **2012**, 109, 026101.

[56]   L. J. Lauhon, W. Ho, *Phys. Rev. Lett.* **2000**, 84, 1527.

[57]   S.-W. Hla, L. Bartels, G. Meyer, K.-H. Rieder, *Phys. Rev. Lett.* **2000**, 85, 2777.

[58]   A. Zhao, Q. Li, L. Chen, H. Xiang, W. Wang, S. Pan, B. Wang, X. Xiao, J. Yang, J. G. Hou, Q. Zhu, *Science* **2005**, 309, 1542.

[59]   N. Pavliček, B. Schuler, S. Collazos, N. Moll, D. Pérez, E. Guitián, G. Meyer, D. Peña, L. Gross, *Nat. Chem.* **2015**, 7, 623.

[60]   N. Pavliček, A. Mistry, Z. Majzik, N. Moll, G. Meyer, D. J. Fox, L. Gross, *Nat. Nanotechnol.* **2017**, 12, 308.

[61]   B. Schuler, S. Fatayer, F. Mohn, N. Moll, N. Pavliček, G. Meyer, D. Peña, L. Gross, *Nat. Chem.* **2016**, 8, 220.

[62]   S. Tan, Y. Ji, Y. Zhao, A. Zhao, B. Wang, J. Yang, J. G. Hou, *J. Am. Chem. Soc.* **2011**, 133, 2002.

[63]   S. Tan, Y. Zhao, J. Zhao, Z. Wang, C. Ma, A. Zhao, B. Wang, Y. Luo, J. Yang, J. Hou, *Phys. Rev. B* **2011**, 84, 155418.

[64]   S. Tan, H. Feng, Y. Ji, Y. Wang, J. Zhao, A. Zhao, B. Wang, Y. Luo, J. Yang, J. G. Hou, *J. Am. Chem. Soc.* **2012**, 134, 9978.

[65]   S. Tan, H. Feng, Y. Ji, Q. Zheng, Y. Shi, J. Zhao, A. Zhao, J. Yang, Y. Luo, B. Wang, J. G. Hou, *J. Phys. Chem. C* **2018**, 122, 28805.

[66]   a) J. Cai, P. Ruffieux, R. Jaafar, M. Bieri, T. Braun, S. Blankenburg, M. Muoth, A. P. Seitsonen, M. Saleh, X. Feng, K. Müllen, R. Fasel, *Nature* **2010**, 466, 470; b) C. Ma, L. Liang, Z. Xiao, A. A. Puretzky, K. Hong, W. Lu, V. Meunier, J. Bernholc, A.-P. Li, *Nano Lett.* **2017**, 17, 6241; c) C. Ma, Z. Xiao, A. A. Puretzky, A. P. Baddorf, W. Lu, K. Hong, J. Bernholc, A.-P. Li, *Phys. Rev. Mater.* **2018**, 2, 014006.

[67]   a) R. B. Woodward, R. Hoffmann, *J. Am. Chem. Soc.* **1965**, 87, 395; b) R. B. Woodward, R. Hoffmann, *Angew. Chem. Int. Ed.* **1969**, 8, 781.

[68]   M. Grzybowski, K. Skonieczny, H. Butenschön, D. T. Gryko, *Angew. Chem. Int. Ed.* **2013**, 52, 9900.

[69]   a) P. Rempala, J. Kroulík, B. T. King, *J. Am. Chem. Soc.* **2004**, 126, 15002; b) L. Zhai, R. Shukla, S. H. Wadumethrige, R. Rathore, *J. Org. Chem.* **2010**, 75, 4748.

[70]   a) T. H. Vo, M. Shekhirev, D. A. Kunkel, M. D. Morton, E. Berglund, L. Kong, P. M. Wilson, P. A. Dowben, A. Enders, A. Sinitskii, *Nat. Commun.* **2014**, 5, 3189; b) A. Narita, X. Feng, Y. Hernandez, S. A. Jensen, M. Bonn, H. Yang, I. A. Verzhbitskiy, C. Casiraghi, M. R. Hansen, A. H. R. Koch, G. Fytas, O. Ivasenko, B. Li, K. S. Mali, T. Balandina, S. Mahesh, S. De Feyter, K. Müllen, *Nat. Chem.* **2014**, 6, 126.

[71]   Z. Xiao, C. Ma, J. Huang, L. Liang, W. Lu, K. Hong, B. G. Sumpter, A.-P. Li, J. Bernholc, *Adv. Theory Simul.* **2019**, 2, 1800172.

[72]   S. Blankenburg, J. Cai, P. Ruffieux, R. Jaafar, D. Passerone, X. Feng, K. Müllen, R. Fasel, C. A. Pignedoli, *ACS Nano* **2012**, 6, 2020.

[73]   a) J. Cai, C. A. Pignedoli, L. Talirz, P. Ruffieux, H. Söde, L. Liang, V. Meunier, R. Berger, R. Li, X. Feng, K. Müllen, R. Fasel, *Nat. Nanotechnol.* **2014**, 9, 896; b) Y.-C. Chen, T. Cao, C. Chen, Z. Pedramrazi, D. Haberer, D. G. de Oteyza, F. R. Fischer, S. G. Louie, M. F. Crommie, *Nat. Nanotechnol.* **2015**, 10, 156.

[74]   P. H. Jacobse, M. J. J. Mangnus, S. J. M. Zevenhuizen, I. Swart, *ACS Nano* **2018**, 12, 7048.

[75]   P. H. Jacobse, A. Kimouche, T. Gebraad, M. M. Ervasti, J. M. Thijssen, P. Liljeroth, I. Swart, *Nat. Commun.* **2017**, 8, 119.

[76]   a) P. B. Bennett, Z. Pedramrazi, A. Madani, Y.-C. Chen, D. G. de Oteyza, C. Chen, F. R. Fischer, M. F. Crommie, J. Bokor, *Appl. Phys. Lett.* **2013**, 103, 253114; b) J. P. Llinas, A. Fairbrother, G. Borin Barin, W. Shi, K. Lee, S. Wu, B. Yong Choi, R. Braganza, J.

Lear, N. Kau, W. Choi, C. Chen, Z. Pedramrazi, T. Dumslaff, A. Narita, X. Feng, K. Müllen, F. Fischer, A. Zettl, P. Ruffieux, E. Yablonovitch, M. Crommie, R. Fasel, J. Bokor, *Nat. Commun.* **2017**, 8, 633.

[77]   L. Yang, C.-H. Park, Y.-W. Son, M. L. Cohen, S. G. Louie, *Phys. Rev. Lett.* **2007**, 99, 186801.

[78]   a) A. Kimouche, M. M. Ervasti, R. Drost, S. Halonen, A. Harju, P. M. Joensuu, J. Sainio, P. Liljeroth, *Nat. Commun.* **2015**, 6, 10177; b) Y.-W. Son, M. L. Cohen, S. G. Louie, *Phys. Rev. Lett.* **2006**, 97, 216803.

[79]   a) R. A. Wolkow, L. Livadaru, J. Pitters, M. Taucerg, P. Piva, M. Salomons, M. Cloutier, B. Martins, in *Field-Coupled Nanocomputing* (Eds: N. Anderson, S. Bhanja), Springer, Berlin, Heidelberg **2014**, p. 33; b) M. Kolmer, C. Joachim, *On-Surface Atomic Wires and Logic Gates*, Springer International Publishing, **2017**; c) T. Skeren, N. Pascher, A. Garnier, P. Reynaud, E. Rolland, A. Thuaire, D. Widmer, X. Jehl, A. Fuhrer, *Nanotechnology* **2018**, 29, 435302.

[80]   M. Kolmer, S. Godlewski, H. Kawai, B. Such, F. Krok, M. Saeys, C. Joachim, M. Szymonski, *Phys. Rev. B* **2012**, 86.

[81]   R. Achal, M. Rashidi, J. Croshaw, D. Churchill, M. Taucer, T. Huff, M. Cloutier, J. Pitters, R. A. Wolkow, *Nat. Commun.* **2018**, 9, 2778.

[82]   a) R. M. Tromp, R. J. Hamers, J. E. Demuth, *Phys. Rev. Lett.* **1985**, 55, 1303; b) R. A. Wolkow, *Phys. Rev. Lett.* **1992**, 68, 2636; c) H. J. W. Zandvliet, *Phys. Rep.* **2003**, 388, 1.

[83]   a) J. J. Boland, *Phys. Rev. B* **1991**, 44, 1383; b) J. W. Lyding, T. C. Shen, J. S. Hubacek, J. R. Tucker, G. C. Abeln, *Appl. Phys. Lett.* **1994**, 64, 2010; c) G. P. Lopinski, D. J. Moffatt, D. D. Wayner, R. A. Wolkow, *Nature* **1998**, 392, 909; d) P. G. Piva, G. A. DiLabio, J. L. Pitters, J. Zikovsky, M. Rezeq, S. Dogel, W. A. Hofer, R. A. Wolkow, *Nature* **2005**, 435, 658.

[84]   a) J. J. Boland, *Phys. Rev. Lett.* **1991**, 67, 1539; b) T. C. Shen, C. Wang, G. C. Abeln, J. R. Tucker, J. W. Lyding, P. Avouris, R. E. Walkup, *Science* **1995**, 268, 1590; c) E. T. Foley, A. F. Kam, J. W. Lyding, P. Avouris, *Phys. Rev. Lett.* **1998**, 80, 1336; d) J. Y. Maeng, J. Y. Lee, Y. E. Cho, S. Kim, S. K. Jo, *Appl. Phys. Lett.* **2002**, 81, 3555.

[85]   G. Scappucci, G. Capellini, W. C. T. Lee, M. Y. Simmons, *Nanotechnology* **2009**, 20, 495302.

[86]   a) M. Kolmer, S. Godlewski, R. Zuzak, M. Wojtaszek, C. Rauer, A. Thuaire, J. M. Hartmann, H. Moriceau, C. Joachim, M. Szymonski, *Appl. Surf. Sci.* **2014**, 288, 83; b) J. Wyrick, X. Q. Wang, P. Namboodiri, S. W. Schmucker, R. V. Kashid, R. M. Silver, *Nano Lett.* **2018**, 18, 7502.

[87]   M. Kolmer, S. Godlewski, J. Lis, B. Such, L. Kantorovich, M. Szymonski, *Microelectron. Eng.* **2013**, 109, 262.

[88]   K. Tomatsu, K. Nakatsuji, T. Iimori, Y. Takagi, H. Kusuhara, A. Ishii, F. Komori, *Science* **2007**, 315, 1696.

[89]   B. Naydenov, I. Rungger, M. Mantega, S. Sanvito, J. J. Boland, *Nano Lett.* **2015**, 15, 2881.

[90]   a) K. Sagisaka, D. Fujita, *Phys. Rev. B* **2005**, 72, 235327; b) K. Nakatsuji, Y. Takagi, F. Komori, H. Kusuhara, A. Ishii, *Phys. Rev. B* **2005**, 72, 241308; c) K. Sagisaka, D. Fujita, *Appl. Phys. Lett.* **2006**, 88, 203118.

[91]   Y. Takagi, K. Nakatsuji, Y. Yoshimoto, F. Komori, *Phys. Rev. B* **2007**, 75, 115304.

[92]   a) A. Bellec, F. Ample, D. Riedel, G. Dujardin, C. Joachim, *Nano Lett.* **2009**, 9, 144; b) S. Godlewski, M. Kolmer, H. Kawai, B. Such, R. Zuzak, M. Saeys, P. de Mendoza, A. M. Echavarren, C. Joachim, M. Szymonski, *ACS Nano* **2013**, 7, 10105; c) A. Radocea, T. Sun, T. H. Vo, A. Sinitskii, N. R. Aluru, J. W. Lyding, *Nano Lett.* **2017**, 17, 170; d)

F. Eisenhut, J. Kruger, D. Skidin, S. Nikipar, J. M. Alonso, E. Guitian, D. Perez, D. A. Ryndyk, D. Pena, F. Moresco, G. Cuniberti, *Nanoscale* **2018**, 10, 12582.

[93]   M. C. Hersam, N. P. Guisinger, J. W. Lyding, *Nanotechnology* **2000**, 11, 70.

[94]   J. B. Ballard, J. H. G. Owen, W. Owen, J. R. Alexander, E. Fuchs, J. N. Randall, J. R. Von Ehr, S. McDonnell, D. D. Dick, R. M. Wallace, Y. J. Chabal, M. R. Bischof, D. L. Jaeger, R. F. Reidy, J. Fu, P. Namboodiri, K. Li, R. M. Silver, *J. Vac. Sci. Technol. B* **2014**, 32, 041804.

[95]   a) A. Fuhrer, M. Fuchsle, T. C. G. Reusch, B. Weber, M. Y. Simmons, *Nano Lett.* **2009**, 9, 707; b) G. Scappucci, G. Capellini, B. Johnston, W. M. Klesse, J. A. Miwa, M. Y. Simmons, *Nano Lett.* **2011**, 11, 2272; c) G. Scappucci, G. Capellini, W. M. Klesse, M. Y. Simmons, *Nanoscale* **2013**, 5, 2600.

[96]   a) S. Godlewski, M. Engelund, D. Pena, R. Zuzak, H. Kawai, M. Kolmer, J. Caeiro, E. Guitian, K. P. C. Vollhardt, D. Sanchez-Portal, M. Szymonski, D. Perez, *Phys. Chem. Chem. Phys.* **2018**, 20, 11037; b) S. Godlewski, H. Kawai, M. Engelund, M. Kolmer, R. Zuzak, A. Garcia-Lekue, G. Novell-Leruth, A. M. Echavarren, D. Sanchez-Portal, C. Joachim, M. Saeys, *Phys. Chem. Chem. Phys.* **2016**, 18, 16757; c) S. Godlewski, M. Kolmer, M. Engelund, H. Kawai, R. Zuzak, A. Garcia-Lekue, M. Saeys, A. M. Echavarren, C. Joachim, D. Sanchez-Portal, M. Szymonski, *Phys. Chem. Chem. Phys.* **2016**, 18, 3854.

[97]   a) S. R. Schofield, P. Studer, C. F. Hirjibehedin, N. J. Curson, G. Aeppli, D. R. Bowler, *Nat. Commun.* **2013**, 4, 1649; b) M. Engelund, R. Zuzak, S. Godlewski, M. Kolmer, T. Frederiksen, A. Garcia-Lekue, D. Sanchez-Portal, M. Szymonski, *Sci. Rep.* **2015**, 5, 14496; c) M. Kolmer, R. Zuzak, G. Dridi, S. Godlewski, C. Joachim, M. Szymonski, *Nanoscale* **2015**, 7, 12325; d) M. Rashidi, M. Taucer, I. Ozfidan, E. Lloyd, M. Koleini, H. Labidi, J. L. Pitters, J. Maciejko, R. A. Wolkow, *Phys. Rev. Lett.* **2016**, 117, 276805; e) M. Yengui, E. Duverger, P. Sonnet, D. Riedel, *Nat. Commun.* **2017**, 8; f) M. Rashidi, W. Vine, T. Dienel, L. Livaduru, J. Retallick, T. Huff, K. Walus, R. A. Wolkow, *Phys. Rev. Lett.* **2018**, 121, 166801; g) T. Huff, H. Labidi, M. Rashidi, L. Livaduru, T. Dienel, R. Achal, W. Vine, J. Pitters, R. A. Wolkow, *Nat. Electron.* **2018**, 1, 636; h) M. B. Haider, J. L. Pitters, G. A. DiLabio, L. Livaduru, J. Y. Mutus, R. A. Wolkow, *Phys. Rev. Lett.* **2009**, 102, 046805.

[98]   S. W. Schmucker, N. Kumar, J. R. Abelson, S. R. Daly, G. S. Girolami, M. R. Bischof, D. L. Jaeger, R. F. Reidy, B. P. Gorman, J. Alexander, J. B. Ballard, J. N. Randall, J. W. Lyding, *Nat. Commun.* **2012**, 3, 935.

[99]   S. Chen, H. Xu, K. E. J. Goh, L. Liu, J. N. Randall, *Nanotechnology* **2012**, 23, 275301.

[100]  M. Moller, S. P. Jarvis, L. Guerinet, P. Sharp, R. Woolley, P. Rahe, P. Moriarty, *Nanotechnology* **2017**, 28, 075302.

[101]  a) N. Pavlicek, Z. Majzik, G. Meyer, L. Gross, *Appl. Phys. Lett.* **2017**, 111, 053104; b) T. R. Huff, H. Labidi, M. Rashidi, M. Koleini, R. Achal, M. H. Salomons, R. A. Wolkow, *ACS Nano* **2017**, 11, 8636.

[102]  G. Dridi, O. F. Namarvar, C. Joachim, *Quantum Sci. Technol.* **2018**, 3, 025005.

[103]  M. Taucer, L. Livaduru, P. G. Piva, R. Achal, H. Labidi, J. L. Pitters, R. A. Wolkow, *Phys. Rev. Lett.* **2014**, 112, 256801.

[104]  J. L. Pitters, I. A. Dogel, R. A. Wolkow, *ACS Nano* **2011**, 5, 1984.

[105]  R. J. Celotta, S. B. Balakirsky, A. P. Fein, F. M. Hess, G. M. Rutter, J. A. Stroscio, *Rev. Sci. Instrum.* **2014**, 85, 121301.

[106]  R. K. Vasudevan, M. Ziatdinov, S. Jesse, S. V. Kalinin, *Nano Lett.* **2016**, 16, 5574.

[107]  M. Koch, J. G. Keizer, P. Pakkiam, D. Keith, M. G. House, E. Peretz, M. Y. Simmons, *Nat. Nanotechnol.* **2019**, 14, 137.

[108]  F. J. Ruess, L. Oberbeck, M. Y. Simmons, K. E. J. Goh, A. R. Hamilton, T. Hallam, S. R. Schofield, N. J. Curson, R. G. Clark, *Nano Lett.* **2004**, 4, 1969.